\documentclass[pdflatex, sn-nature]{sn-jnl}% Style for submissions to Nature Portfolio journals
\usepackage{graphicx}%
\usepackage{multirow}%
\usepackage{amsmath,amssymb,amsfonts}%
\usepackage{amsthm}%
\usepackage{mathrsfs}%
\usepackage[title]{appendix}%
\usepackage{xcolor}%
\usepackage{textcomp}%
\usepackage{manyfoot}%
\usepackage{booktabs}%
\usepackage{algorithm}%
\usepackage{algorithmicx}%
\usepackage{algpseudocode}%
\usepackage{listings}%
\usepackage{empheq}
\usepackage{lmodern}
\usepackage{mathrsfs}
\usepackage{fix-cm}
\usepackage{scalefnt}

\begin{document}

\title[Non-Hermitian Topology of a Subwavelength Resonator in Synthetic Space]{Non-Hermitian Topology of a Subwavelength Resonator in Synthetic Space}


\author[1,2]{\fnm{N.} \sur{Solodovchenko}$^{\dagger\text{,}}$}

\author[1,2]{\fnm{F.} \sur{Zhang}$^{\dagger\text{,}}$}

\author[2]{\fnm{M.} \sur{Bochkarev}}

\author[2]{\fnm{K.} \sur{Samusev}}

\author*[1]{\fnm{M.} \sur{Song}}\email{kevinsmz@foxmail.com}

\author*[2]{\fnm{M.} \sur{Limonov}} \email{m.limonov@mail.ioffe.ru}

\author*[1,2]{\fnm{A.} \sur{Bogdanov}}\email{a.bogdanov@hrbu.edu.cn}

\affil[1]{\orgdiv{Qingdao Innovation and Development Center}, \orgname{Harbin Engineering University}, \orgaddress{\street{Qingdao}, \city{Shandong}, \postcode{266000}, \country{China}}}

\affil[2]{\orgdiv{School of Physics and Engineering}, \orgname{ITMO University}, \orgaddress{\street{Saint-Petersburg}, \postcode{197101}, \country{Russia}}}

%\affil[3]{\orgdiv{Ioffe Institute}, \orgname{St. Petersburg}, \orgaddress{\street{St. Petersburg}, \postcode{194021}, \country{Russia}}}

\affil[]{\textsuperscript{$\dagger$}These authors contributed equally to this work}

%% =================================NEW ABSTRACT
\abstract{Non-Hermitian singularities and the nontrivial topology of energy bands have been extensively explored in extended periodic systems, where phenomena such as exceptional points and bulk Fermi arcs are naturally associated with periodicity and momentum space. Here, we show that this non-Hermitian topology can be extended to the domain of subwavelength resonators using a synthetic space. By reconstructing the complex energy bands of a subwavelength ring resonator in the synthetic space of geometrical parameters, we experimentally observe a spectral triplet formed by two exceptional points (EPs) connected by a bulk Fermi arc and a topologically robust quasi-bound state in the continuum (q-BIC). This Fermi-arc q-BIC emerges exactly at the intersection of the bulk Fermi arc and the q-BIC line in synthetic space, establishing a direct topological link between two spectral singularities. Encircling the EPs in synthetic space visualizes the nontrivial topology of the complex energy-band structure and reveals the opposite half-integer topological charges of the EPs connected by the bulk Fermi arc. These results extend non-Hermitian topology to subwavelength structures and establish compact resonators as a versatile platform for applications in photonics, microwave physics, and acoustics.}




\maketitle

\newpage

Topology provides a general framework for relating global spectral structure to robust physical observables and functionalities~\cite{Lu2016Jul}. In photonics, this framework has been fundamentally enriched by non-Hermitian systems, whose spectra can host topological features that have no counterpart in closed Hermitian systems~\cite{Midya2018Jul,Ozawa2019Mar,Kawabata2019Aug,Bergholtz2021Feb,Parto2021Jan,Okuma2023Mar,Wang2023Jun,Kozii2024Jun}. Among the most characteristic examples are exceptional points (EPs), at which two or more eigenstates coalesce, become self-orthogonal, and share the same complex eigenvalue, in contrast to diabolic points, where the eigenstates remain orthogonal~\cite{Miri2019Jan,zhang2026exceptionalpointsphotonicsnonhermitian,Kozii2024Jun,Zhou2018Mar,Wang2019Feb,Yang2018Jan,Hashemi2022Jun}. The mathematical foundations of such singularities trace back to Kato’s treatment of linear operators with parameter-dependent spectra~\cite{kato2013perturbation}. Beyond isolated EPs, non-Hermitian topology includes higher-order EPs, exceptional lines, rings, and surfaces~\cite{Mandal2021Oct,Moors2019Jan,Zhen2015Sep,Zhou2019Feb,Zhang2019Dec,Ding2018Aug}, together with braiding of exceptional singularities and the associated non-Abelian phenomena~\cite{dingNonHermitianTopologyExceptionalpoint2022,guoExceptionalNonAbelianTopology2023,yangNonAbelianPhysicsLight2024}.

A particularly striking manifestation of the transition from Hermitian to non-Hermitian topology is the emergence of a bulk Fermi arc connecting EPs in the momentum space. When loss perturbs a Hermitian degeneracy, a Dirac point with nontrivial Berry phase splits into a pair of EPs with opposite half-integer topological charges~\cite{pujol2023dirac,Zhou2018Mar}. The corresponding complex spectrum acquires a two-sheeted Riemann-surface structure, in which the EPs serve as branch-point singularities and the bulk Fermi arc marks the locus where one component of the complex eigenfrequency remains degenerate~\cite{Kozii2024Jun,Zhou2018Mar}. Although EPs have been realized in many non-Hermitian platforms~\cite{dembowskiExperimentalObservationTopological2001,dingEmergenceCoalescenceTopological2016,leeObservationExceptionalPoint2009,chenRevealingMissingDimension2020}, bulk Fermi arcs remain much more naturally associated with spatial or angular periodicity and an underlying momentum space~\cite{Tlusty2021Aug, zhenSpawningRingsExceptional2015,pujol-closaDiracPointsEmbedded2023,yang2023realization}. Despite multiple theoretical studies of the bulk Fermi arc in various systems like superconductors, heavy-fermion systems, and disordered Dirac materials~\cite{Cayao2023Mar,Nagai2020Nov,Zhao2018Aug,pengLossinducedSuppressionRevival2014}, its experimental observation has so far been restricted to a two-dimensional photonic crystal slab~\cite{Zhou2018Mar} and to an orbital-angular-momentum lattice~\cite{yang2023realization}.

Although non-Hermitian phenomena are most commonly discussed in momentum space, they can emerge more generally in {\it synthetic spaces} formed by tunable system parameters~\cite{yuan2018synthetic}. In photonics, synthetic dimensions provide an alternative route to non-Hermitian topology by promoting controllable degrees of freedom—such as frequency, orbital angular momentum, or photon number—to effective dimensions, thereby enabling topological structures and wave dynamics beyond the constraints of real space, as recently demonstrated for optical propagation and interference in photon-number Fock space~\cite{arguello2024synthetic,ehrhardt2023perspective,xu2026principles}. Within this framework, EPs and related non-Hermitian effects have been realized in synthetic spaces~\cite{zhang2019experimental,dingEmergenceCoalescenceTopological2016}, demonstrating the broad flexibility of the synthetic-space approach. Despite large interest in non-Hermitian systems and synthetic spaces, the complex energy-band Riemann surface remained experimentally inaccessible until very recently, when synthetic frequency dimensions in a modulated photonic ring resonator enabled its reconstruction and revealed branch points, branch cuts, and complex-energy winding as fundamental signatures of non-Hermitian band topology~\cite{cheng2026experimental}. In parallel, subwavelength dielectric resonators have become a key platform of modern photonics owing to their strong field confinement, pronounced radiative openness, and ability to support high-$Q$ resonances in compact volumes~\cite{kivshar2022rise, wonMietronicEra2019a}. Of particular importance is their ability to host high-$Q$ quasi-bound states in the continuum (q-BICs)~\cite{koshelev2023bound,Rybin2017Dec,Bogdanov2019Jan,Mylnikov2020Jun,Odit2021Jan}, which have enabled enhanced nonlinear response~\cite{Koshelev2020Jan,Zalogina2023Apr}, lasing~\cite{Tiguntseva2020Jul,Mylnikov2020Jun,zhang2018phonon}, and sensing~\cite{Jacobsen2022Jun,Yusupov2021Nov}. The combination of synthetic-space engineering and advances in subwavelength dielectric resonators presents a natural opportunity to extend non-Hermitian physics beyond periodic structures. The remaining open question is whether non-Hermitian topology can survive in a single subwavelength resonator, thereby bringing subwavelength resonators into the broader framework of non-Hermitian physics.

Here, we investigate a single open dielectric ring resonator operating in the subwavelength regime, with a characteristic size-to-wavelength ratio $\lambda/L \sim 3$, in a two-dimensional synthetic space defined by its geometrical parameters. By measuring extinction spectra in the GHz range and applying the quasinormal-mode (QNM) formalism, we fully reconstruct the complex energy band surfaces. The reconstructed spectrum reveals a triplet of spectral singularities formed by two EPs connected by a bulk Fermi arc and a quasi-bound state in the continuum (q-BIC) emerging at the intersection of the Fermi arc with the q-BIC line. We refer to this topologically robust state as a {\it Fermi-arc q-BIC}, emphasizing that its radiation suppression is embedded in the global non-Hermitian topology of the spectrum. Encircling the EPs in synthetic space resolves their opposite half-integer topological charges, distinguishing the nontrivial loop around a single EP from the topologically trivial loop enclosing both EPs. These results show that the non-Hermitian topology associated with bulk Fermi arcs can be extended to a single subwavelength ring resonator.

{\bf Exceptional points and bulk Fermi arc in subwavelength resonators}. Phenomenologically, the interaction between two modes in an open system can be described by an effective non-Hermitian Hamiltonian~\cite{Hsu2016Jul}:
\begin{equation}
\mathcal{\hat H}=\left(\begin{array}{cc}
\omega_{1} & \varkappa  \\
\varkappa & \omega_{2}
\end{array}\right)-i\left(\begin{array}{cc}
\gamma_{1} & \sqrt{\gamma_{1} \gamma_{2}} \\
\sqrt{\gamma_{1} \gamma_{2}} & \gamma_{2}
\end{array}\right),
\label{eq1}
\end{equation}
where $\omega_{1,2}$ and $\gamma_{1,2}$ are the resonant frequencies and radiative losses of the uncoupled resonant states, respectively; $\varkappa$ is the near-field coupling between the states, while $\sqrt{\gamma_1\gamma_2}$ accounts for the coupling through the radiation continuum. Within the two-mode approximation, the BIC is formed when the coupling coefficient $\varkappa$ has the following form~\cite{volya2003non}:
\begin{equation}
\varkappa_{\mathrm{BIC}}=\Delta\omega\frac{\sqrt{\gamma_{1} \gamma_{2}}}{\left(\gamma_{1}-\gamma_{2}\right)}. \label{eq2}
\end{equation}

While for any fixed $\varkappa_\text{BIC}$ and unequal losses $\gamma_1 \neq \gamma_2$, a BIC can be achieved by varying $\Delta\omega=\omega_1 - \omega_2$, the EP can be reached only for certain coupling parameters and frequency detuning~\cite{Solodovchenko2024Feb}:
\begin{subequations}
    \begin{empheq}[left={\empheqlbrace\,}]{align}
        &\varkappa_\text{EP}= \pm \dfrac{\left|\gamma_{1}-\gamma
        _{2}\right|}{2}, \label{eq3a}  \\
        & \Delta\omega = \mp 2 \frac{\sqrt{\gamma_{1} \gamma_{2}}}{\mathrm{sgn} (\gamma_1-\gamma_2)}. \label{eq3b}
     \end{empheq}
\end{subequations}
It is important to note that according to Eq.~\eqref{eq3a}, the distance between two EPs can be changed by varying $|\gamma_1-\gamma_2|$ and, moreover, transform to the Dirac point at $\gamma_1 = \gamma_2$ (see Supplementary Information, Sec.~I).

The bulk Fermi arc is the locus in parameter space along which one component of the complex eigenfrequency becomes degenerate. In the present case, it is given by:
\begin{equation}
    \varkappa_\text{FA}=-\Delta\omega\dfrac{\left(\gamma_{1}-\gamma_{2}\right)}{4 \sqrt{\gamma_{1} \gamma_{2}}}. \label{eq4}
\end{equation}

\begin{figure}[t]
\centering
\includegraphics[width=0.8\textwidth]{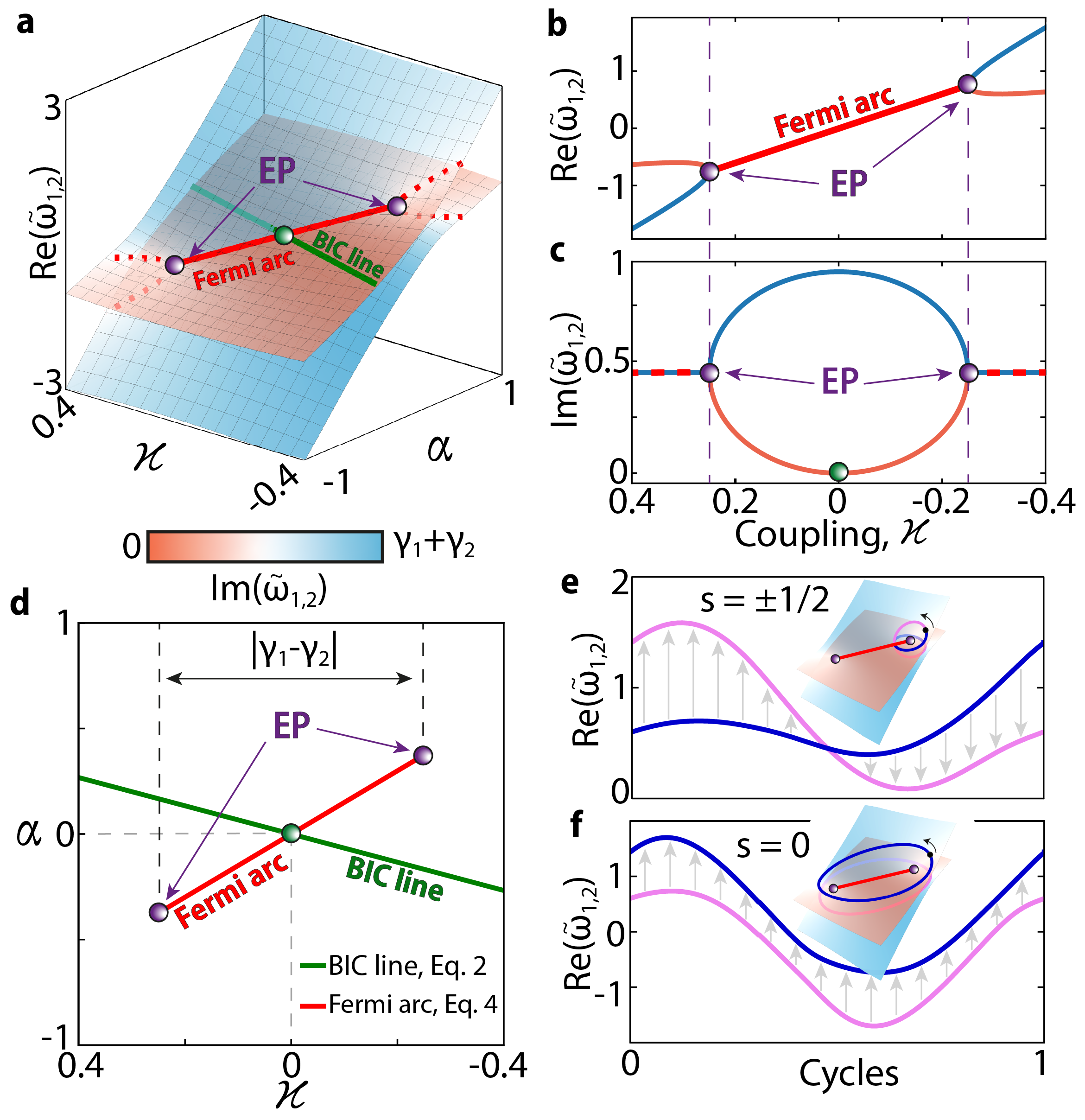}
\caption{$\mid$ \textbf{Exceptional points and bulk Fermi in the synthetic space within the effective Hamiltonian approach.} \textbf{a}, Two-sheeted Riemann surface of the complex eigenfrequencies $\widetilde{\omega}_{1,2}$ in the synthetic space $(\varkappa,\alpha)$, with the real part shown along the vertical axis and the imaginary part encoded by color. \textbf{b,c}, Real and imaginary parts of $\widetilde{\omega}_{1,2}$ as functions of the coupling parameter $\varkappa$ along the bulk Fermi arc. The purple dotted lines mark the positions of the EPs. All results are obtained from the phenomenological effective Hamiltonian~\eqref{eq1}. \textbf{d}, Intersection of the bulk Fermi arc and the BIC line defined by Eqs.~\eqref{eq2} and~\eqref{eq4}. The distance between the two EPs is $|\gamma_{1}-\gamma_{2}|$, and the topologically robust BIC is indicated by the green point. \textbf{e,f}, Evolution of the eigenfrequencies along closed paths on the Riemann surface encircling a single EP and both EPs, respectively.}\label{fig1}
\end{figure}

Figure~\ref{fig1}a shows the eigenfrequency spectrum of Hamiltonian~\eqref{eq1} assuming the internal coupling $\varkappa$ between the modes and the initial frequency detuning $\Delta\omega= 2\alpha$ as independently varying parameters of the synthetic space. The loss rates, for instance, are fixed as $\gamma_{1} = 0.7$ and $\gamma_{2} = 0.2$.  The spectrum represents a two-sheeted Riemann surface with two EPs connected by a bulk Fermi arc. The EPs appear under conditions~\eqref{eq3a} and~\eqref{eq3b}. The bulk Fermi arc spans the region of the synthetic space between the two EPs, where the real parts of the eigenfrequencies coincide while the imaginary parts repulse. The bulk Fermi arc, which is defined by Eq.~\eqref{eq4}, is visualized in Fig.~\ref{fig1}a. As shown in Figs.~\ref{fig1}b and~\ref{fig1}c, along the bulk Fermi arc, the real parts of the eigenfrequencies $\tilde{\omega}_{1,2}$ are degenerate, while their imaginary parts repel, and at $\kappa=0$ the bulk Fermi arc intersects the BIC-line, giving rise to a topologically robust BIC on the arc. Indeed, the conditions for the BIC [Eq.~\eqref{eq2}] and the bulk Fermi arc [Eq.~\eqref{eq4}] define two lines in the synthetic space that intersect at $\varkappa=0$, giving rise to a topologically robust BIC on the bulk Fermi arc (Fig.~\ref{fig1}d). This establishes a fundamental connection between EPs and BICs in non-Hermitian systems, which was recently discussed by the example of a planar slab with hyperbolic dispersion on a birefringent substrate~\cite{pujol2023dirac} and other non-Hermitian platforms~\cite{PhysRevLett.134.103802, wu2026experimentalobservationexceptionalbound}.

Encircling a single EP on the Riemann surface causes the eigenmodes to exchange identities (Fig.~\ref{fig1}e). To recover the initial state, a double $4\pi$ traversal is required, reflecting the presence of an EP with topological charge $s=\pm 1/2$ inside the loop. The two EPs carry opposite charges, as confirmed by a loop enclosing both simultaneously (Fig.~\ref{fig1}f). In this case, a single $2\pi$ traversal is sufficient to return to the starting state, demonstrating that the total topological charge sums to $s=0$.

{\bf Bulk Fermi arc in a subwavelength ring resonator}. To proceed from the simplified effective non-Hermitian Hamiltonian model to a realistic system, we consider a subwavelength ring resonator made of high-index ceramic with $\varepsilon = 80.05$ and dielectric losses $\tan\delta = 3.6 \cdot 10^{-4}$ (See  Methods). The ring resonator is characterized by the outer radius $R_{\mathrm{out}}$, inner radius $R_{\mathrm{in}}$, and height $h$. The sidewall thickness is defined as $W = R_{\mathrm{out}} - R_{\mathrm{in}}$. For generality, we introduce the dimensionless ratios $R_{\mathrm{in}}/R_{\mathrm{out}}$ and $W/h$, which together form a two-dimensional synthetic space. Equivalently to using complex eigenfrequencies, we represent the results in terms of the dimensionless complex wavenumber $kW$.

We limit our analysis to the case of modes with azimuthal index $m = \pm 1$, as they can be selectively excited by a plane wave incident along the resonator axis. The modes of the resonator can be conditionally grouped into two families: (i) radial modes, quantized across the resonator cross-section, and (ii) axial modes, quantized along the vertical direction~\cite{Rybin2017Dec}. Their distinct spectral shifts under variations of $W$ and $h$ allow controlled mode interaction and enable controlled access to EPs~\cite{Solodovchenko2024Feb,Zhang2025Feb}, q-BICs~\cite{Bogdanov2019Jan, Odit2021Jan}, anapole states~\cite{Baranov2018Jul,Yang2018May}, or superscattering regimes~\cite{CanosValero2023Aug}. 

\begin{figure}[ht]
\centering
\includegraphics[width=1\textwidth]{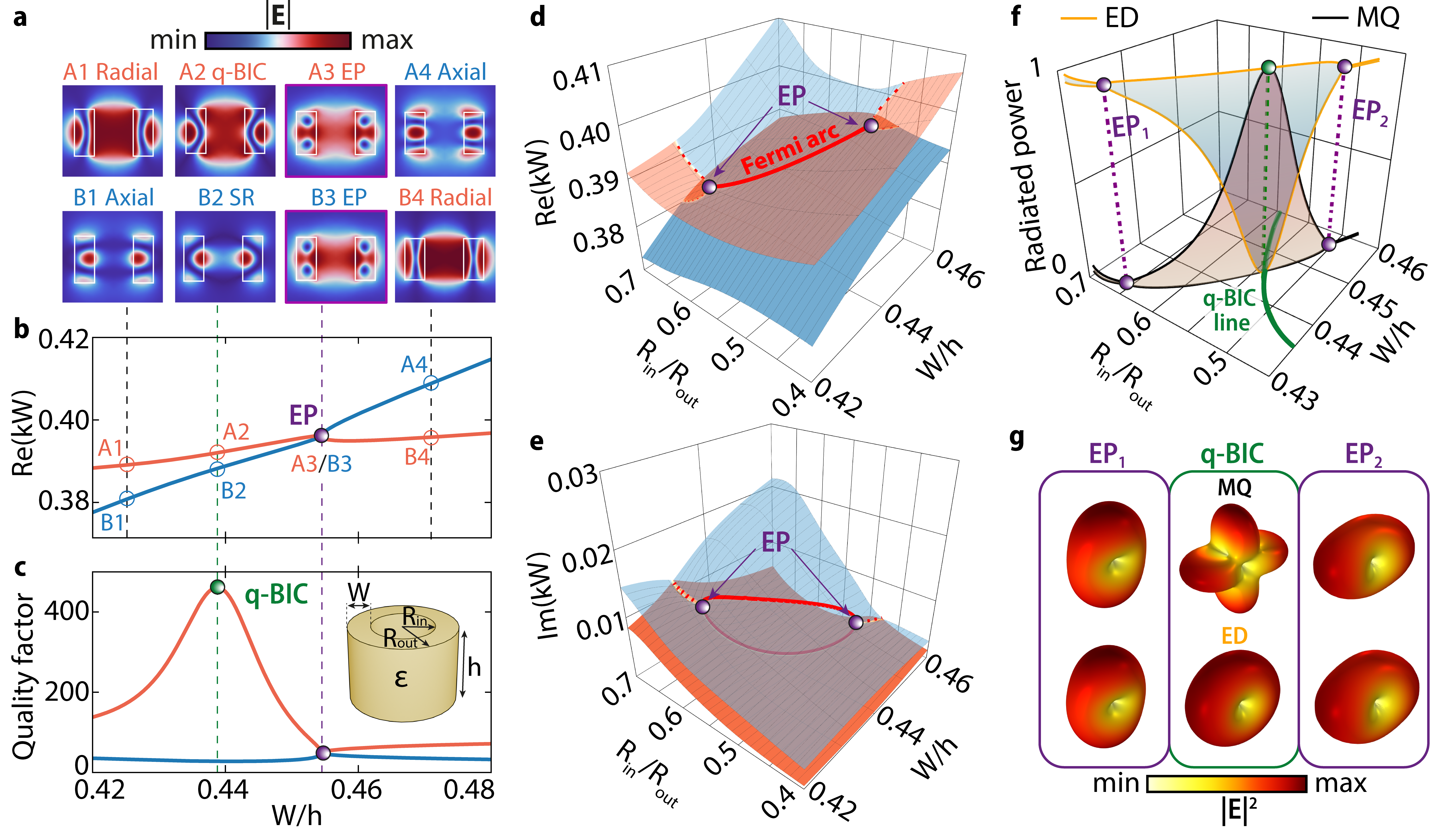}
\caption{$\mid$ {\bf EPs and bulk Fermi arc in the single ring resonator.} {\bf a}, Calculated near field distributions of electric field $|\tilde{\mathbf{E}}|$ in ring resonator for both modes for the corresponding points in panel {\bf b}: at q-BIC, at EP, and at a larger spectral splitting $\Delta \omega$. {\bf b}, Real component of dimensionless eigenfrequency $\operatorname{Re}({kW})$ relative to width-to-height ratio $W/h$. {\bf c}, Dependencies of quality factors of modes on the ratio $W / h$. {\bf b} and {\bf c}, both constructed for a fixed radius ratio $R_{\mathrm{in}} / R_{\mathrm{out}}=0.499$. {\bf d} and {\bf e}, Two-sheet Riemann surface of real and imaginary parts of eigenfrequencies $kW$ in the synthetic space of the resonator. Purple dots mark the positions of EPs, and the red line in panel {\bf d} marks the Fermi arc. {\bf f}, Evolution of the radiated power associated with the electric dipole and magnetic quadrupole channels along the bulk Fermi arc and {\bf g} corresponding radiation patterns for two EPs and the q-BIC state between them.}\label{fig2}
\end{figure}

Figures~\ref{fig2}b and~\ref{fig2}c show the numerically calculated spectral position of Mie resonances and their Q-factors as a function of $W/h$ for the fixed ratio $R_{\mathrm{in}}/R_{\mathrm{out}} = 0.499$. Depending on $W/h$, different regimes can be identified, including a q-BIC at $W/h=0.439$ and an EP at $W/h=0.455$. The corresponding electric field distributions are shown in Fig.~\ref{fig2}a. One can see that at the EP, the two eigenmode profiles become degenerate (A3, B3). To fully uncover the non-Hermitian topology of the spectrum, we map the eigenvalues across the whole synthetic space $(R_{\mathrm{in}}/R_{\mathrm{out}}, W/h)$ of a single resonator. The two-sheet Riemann surfaces of the real and imaginary parts of the eigenfrequencies $kW$ are shown in Figs.~\ref{fig2}d and~\ref{fig2}e, where the purple dots mark the EPs and the red line denotes the bulk Fermi arc connecting them. Along this arc, the real parts of the eigenfrequencies coincide, while the imaginary parts remain distinct.

To gain further insight into the mode evolution along the bulk Fermi arc, we analyze the multipolar decomposition shown in Fig.~\ref{fig2}f. It reveals the interplay between the electric dipole (ED) and magnetic quadrupole (MQ) channels: the q-BIC is characterized by suppression of the ED response, whereas the EP corresponds to the coalescence of two ED-dominated modes. The far-field radiation patterns in Fig.~\ref{fig2}g further illustrate this evolution, showing nearly identical ED-like profiles at the EPs and a suppressed ED contribution with dominant MQ character at the q-BIC. The q-BIC line intersects the bulk Fermi arc (Fig.~\ref{fig2}g), in agreement with the prediction of the effective Hamiltonian~\eqref{eq1}. Thus, despite being only an approximate model of an open photonic system, the effective non-Hermitian Hamiltonian accurately captures all essential spectral and topological features of the ring resonator, including the EPs, the bulk Fermi arc, and the q-BIC residing on it.

{\bf Experimental observation of the bulk Fermi arc}. 
The EP explicitly appears in the eigenvalue spectra, which can be calculated numerically. However, identifying an EP from scattering spectra is challenging as it does not manifests itself explicitly~\cite{duggan2022lim}. Nevertheless, this information can be extracted from the observables. To do that, we analyze the scattering from the resonator under illumination by a linearly polarized plane wave propagating along the rotational symmetry axis (see Supplementary Information, Sec.~IV). The extinction cross-section spectra is given by a superposition of the  Fano resonances (see Eq.~(S12) in the Supplementary Information, Sec.~III), whose parameters can be analytically determined using QNM theory~\cite{Lalanne2018May,Bochkarev2024Dec}. It should be emphasized that the total scattering cross-section is generally not well suited for extracting complex eigenvalues, since several eigenmodes contribute to the same scattering channel. In contrast, the total extinction cross-section $\sigma_\text{ext}$ can be decomposed into partial contributions as $\sigma_{\text{ext}}=\sum_{n} \sigma_\text{ext}^{n}+\sigma_\text{bg}$, where $\sigma_\text{ext}^{n}$ is the partial contribution of the $n$-th QNM and the background term is $\sigma_\text{bg}$ are defined in Sec.~III of the Supplementary Information. This approach enables the extraction of Fano parameters and complex eigenvalues from the measured extinction spectra, allowing us to independently track both modes across the synthetic space $(R_{\mathrm{in}}/R_{\mathrm{out}}, W/h)$ (see Supplementary Information, Sec.~III).

This approach allows us to study independently the behavior of the modal intensities $\sigma_{n}$ and Fano asymmetry parameters $q_{n}$ for both modes in the synthetic space $(R_{\mathrm{in}} / R_{\mathrm{out}}, W / h)$ (see Supplementary Information, Sec.~III). The calculations reveal an interesting feature: near the EP, the intensities of both modes rapidly increase and become almost equal. This phenomenon is associated with the manifestation of self-orthogonality in the close-to-merging regime and is another well-known marker of an EP~\cite{Moiseyev2011Mar} (see Supplementary Information, Sec.~II). This intensity behavior can also be explained by the Petermann factor~\cite{wiersig2023petermann,wang2020petermann}. The Fano parameter $q_n$ provides an additional indicator of q-BIC and anapole states~\cite{Zhang2025Feb} (see Supplementary Information, Sec.~III).

We next present the key experimental results used to reconstruct the bulk Fermi arc and the associated spectral topology. To sample the synthetic parameter space, we fabricated 13 resonator series with fixed values of $R_{\mathrm{in}}/R_{\mathrm{out}}$ and progressively reduced the height of each sample, spanning the range from $W/h=0.42$ to $0.46$ (see Supplementary Information, Sec.~IV). Because the resonator width $W$ varies from series to series, the initial height and polishing step were adjusted accordingly. In total, this procedure yielded 203 extinction spectra across the synthetic parameter space. The experimental extinction cross-sections were measured using the optical theorem~\cite{larsson2013wideband}.

The experimental spectra were fitted using Eq.~S12, with numerically calculated eigenfrequencies, modal intensities, and Fano asymmetry parameters used as initial values in the fitting procedure (see Supplementary Information, Secs.~III and IV). Such theoretically informed fitting is essential near the EP, where the spectral response is strongly distorted and cannot be reliably parameterized without an accurate line-shape model.

The same fitting procedure was then applied to all thirteen resonator series obtained through the height-reduction protocol. For each measured extinction spectrum, the complex eigenfrequencies were reconstructed and subsequently combined into a single dataset in the synthetic space $(R_{\mathrm{in}}/R_{\mathrm{out}},\,W/h)$. This combined reconstruction made it possible to determine both the bulk Fermi arc and the q-BIC line experimentally.

\begin{figure}[ht]
\centering
\includegraphics[width=0.8\textwidth]{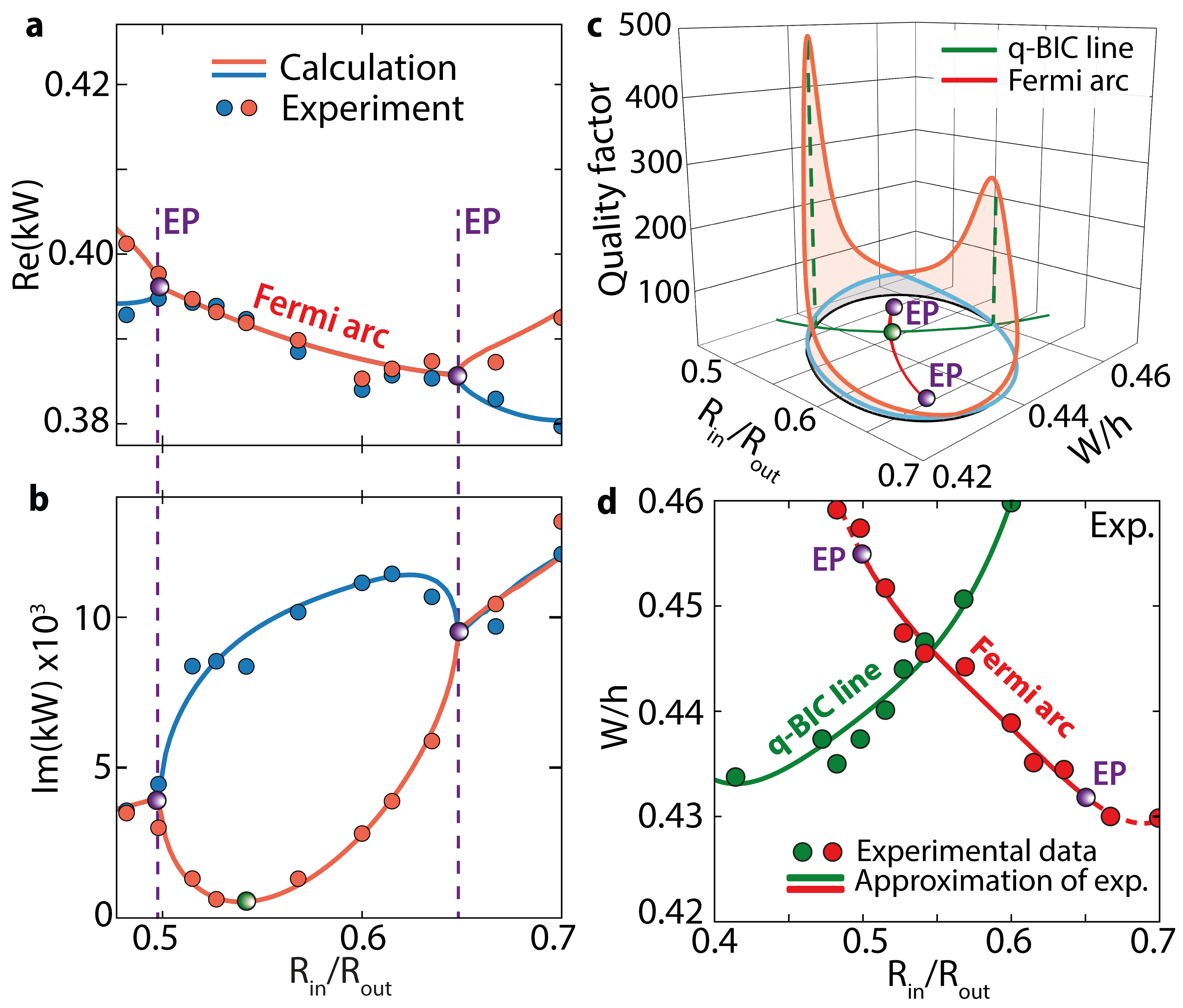}
\caption{$\mid$ \textbf{Experimental observation of the bulk Fermi arc in dielectric ring resonators.} \textbf{a,b}, Real and imaginary parts of the dimensionless eigenfrequencies $kW$ of the two interacting modes traced along the bulk Fermi arc as functions of $R_{\mathrm{in}}/R_{\mathrm{out}}$. Solid curves show numerical calculations, whereas circles denote the values reconstructed from fits of the experimental extinction spectra. The purple dashed lines mark the positions of the exceptional points (EPs). \textbf{c}, Numerically calculated q-BIC line and bulk Fermi arc in the synthetic parameter space $(R_{\mathrm{in}}/R_{\mathrm{out}},\,W/h)$. The two EPs are marked in purple. \textbf{d}, Experimentally reconstructed q-BIC line and bulk Fermi arc obtained from the fitted extinction spectra of all resonator series. The experimental points are shown as circles, and the solid curves represent polynomial approximations. The agreement between \textbf{c} and \textbf{d} demonstrates experimentally the presence of two EPs connected by a bulk Fermi arc and the formation of a topologically robust q-BIC at the intersection of the q-BIC line and the Fermi arc.}\label{fig3}
\end{figure}

The detailed reconstruction of the complex-eigenvalue landscape is presented in Supplementary Information, Sec.~IV. Here, we focus on the main experimental signatures extracted from that analysis. Figures~\ref{fig3}a and~\ref{fig3}b show the real and imaginary parts of the two interacting dimensionless eigenfrequencies $kW$ traced along the bulk Fermi arc as functions of $R_{\mathrm{in}}/R_{\mathrm{out}}$. The circles denote values reconstructed from fits of the experimental extinction spectra, whereas the solid curves show the numerical results. The agreement between experiment and theory is very good over the range where the arc is resolved experimentally. In particular, the two EPs define the ends of the arc, while between them the real parts of the eigenfrequencies remain nearly degenerate and the imaginary parts stay distinct, which is the defining signature of the bulk Fermi arc.

Figures~\ref{fig3}c and~\ref{fig3}d compare the numerically calculated and experimentally reconstructed q-BIC line and bulk Fermi arc in the synthetic parameter space. The experimental data closely follow the numerical trends, demonstrating the presence of two EPs connected by a bulk Fermi arc. Moreover, the intersection of the q-BIC line with the bulk Fermi arc provides direct experimental evidence for a topologically robust q-BIC, in agreement with the prediction of the effective Hamiltonian shown in Fig.~\ref{fig1}d.

\begin{figure}[ht]
\centering
\includegraphics[width=0.8\textwidth]{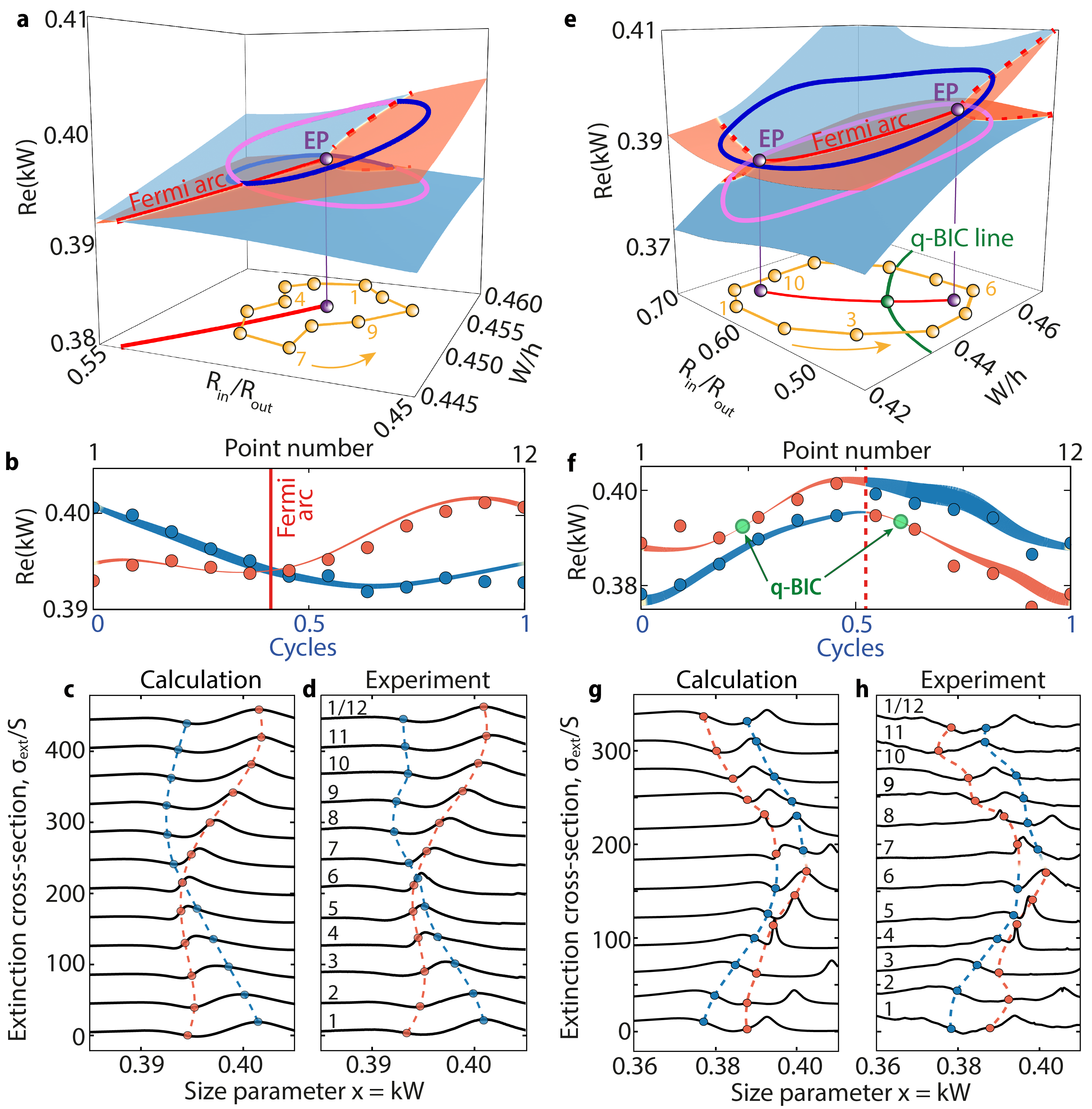}
\caption{$\mid$ \textbf{Transformation of eigenfrequencies and spectra upon encircling exceptional points.} \textbf{a}, Calculated surface of the real part of the eigenfrequencies containing an EP at $R_{\mathrm{in}}/R_{\mathrm{out}}=0.499$ and $W/h=0.455$ (purple dot). The dark blue curve on the surface shows the evolution of the eigenfrequencies around the EP. The corresponding geometrical parameters are indicated by dark blue dots in the $(R_{\mathrm{in}}/R_{\mathrm{out}},\,W/h)$ plane. The numbered light orange dots mark the experimental trajectory. The red dashed line is the projection of the bulk Fermi arc onto the plane. \textbf{b}, Evolution of the real parts of the eigenfrequencies along a single $2\pi$ circuit around the EP, together with the corresponding experimental values plotted as functions of the point number and the polar angle defined in panel \textbf{a}. After one full loop, the two modes exchange identities rather than return to their initial states. \textbf{c,d}, Numerically calculated and experimentally measured extinction spectra, respectively, together with the corresponding eigenfrequencies marked by colored dots. \textbf{e--h}, Same analysis for a closed loop enclosing both EPs. In this case, the loop is topologically trivial, and the eigenmodes do not exchange identities. The line thickness in panels \textbf{b} and \textbf{f} is proportional to the imaginary part of the eigenfrequency.}\label{fig4}
\end{figure}

An EP is a spectral singularity that cannot be accessed directly, even in numerical calculations, but can be identified from the non-trivial topology of the surrounding eigenfrequency surface. To reveal this topology, we trace closed trajectories around the EP in the synthetic space of geometric parameters, both numerically (black curve) and experimentally (light orange curve), as shown in Fig.~\ref{fig4}a. Along such a path, the two Mie resonances exchange identities upon crossing the bulk Fermi arc and move from one Riemann sheet to the other. As shown in Fig.~\ref{fig4}b, after a single $2\pi$ circuit around the EP, the system does not return to its initial state; instead, the two modes are interchanged. Apparently, only after a second circuit, giving a full $4\pi$ cycle, is the initial state recovered. This behavior is directly reflected in the eigenvalue spectra obtained numerically (Fig.~\ref{fig4}c) and experimentally (Fig.~\ref{fig4}d) for the numbered points along the light orange polygonal trajectory in Fig.~\ref{fig4}a. The characteristic $4\pi$ evolution is the spectral analogue of a M\"obius topology and provides a direct signature of the non-Hermitian nature of the EP~\cite{Midya2018Jul,Kawabata2019Aug,Parto2021Jan}.

For comparison, Figs.~\ref{fig4}e–h present the trivial case of encircling both EPs and the bulk Fermi arc simultaneously. In this scenario, the two Mie resonances no longer exchange identities, and a $2\pi$ circulation is sufficient to return to the initial state. This behavior corresponds to a total topological charge of $s=0$, confirming that the two EPs carry opposite charges $s=\pm 1/2$ (see Fig.~\ref{fig1}f). As shown in Fig.~\ref{fig4}f, the trajectory intersects the q-BIC line twice, first on the high-frequency orange branch and then on the low-frequency blue branch (green arrows). The line widths in panels \textbf{b} and \textbf{f} are proportional to the imaginary parts of the eigenfrequencies, highlighting the strong suppression of radiation losses at the q-BICs. Thus, by encircling the EP pair and the bulk Fermi arc, we observe two q-BICs and confirm experimentally that these states form a continuous line in the parametric space that crosses the Fermi arc (see Supplementary Information, Sec. IV).

To conclude, we have shown that non-Hermitian topology can survive in a single subwavelength dielectric resonator. Using a synthetic-space approach, we demonstrated that such a resonator can host a bulk Fermi arc connecting two EPs of opposite topological charge, thereby extending a hallmark feature of non-Hermitian topology beyond periodic structures. We further showed experimentally that the q-BIC emerges as a topologically robust state at the intersection of the bulk Fermi arc and the q-BIC line, thereby establishing a direct topological link between q-BIC, EPs, and bulk Fermi arc. These results bring subwavelength resonators into the broader framework of non-Hermitian physics and establish compact resonators as a versatile platform for applications in photonics, microwave physics, and acoustics. Importantly, these spectral features remain robust against material losses: losses merely shift the positions of the EPs and the bulk Fermi arc in parameter space without destroying them (see Supplementary Information, Sec.~V).

\section*{Online content} 
Any methods, additional references, Nature Portfolio reporting summaries, source data, extended data, supplementary information, acknowledgements, peer review information; details of author contributions and competing interests; and statements of data and code availability
are available at: .....

\bibliography{references}

\section*{Methods}
{\bf Theoretical calculations }
The numerical simulations were performed using COMSOL Multiphysics. For a detailed physical analysis, we applied the QNM theory~\cite{Wei2018QNM, Lalanne2018May,Wu2023Mar}. QNMs are source-free solutions of Maxwell's equations that were obtained using the COMSOL eigensolver and normalized using the PML norm~\cite{Sauvan2022Feb}. The QNM theory allows for the calculation of all spectral response parameters (for more details see Supplementary information, Sec.~III), including the scattering intensities $\sigma_{n}$ and the Fano parameters $q_{n}$~\cite{Bochkarev2024Dec}, thus obtaining 3D $\sigma_{n}$ and $q_{n}$ maps in the parametric space. Each 3D map consists of 301 rows along $R_{\mathrm{in}} / R_{\mathrm{out}}$ in the range from 0.4 to 0.7 and 601 columns along $\left(R_{\mathrm{out}}-R_{\mathrm{in}}\right) / h=W / h$ in the range from 0.42 to 0.46. The numerical extinction cross-section spectra $\sigma_{\mathrm{ext}}$ were obtained as a linear sum of the responses of the single QNM responses under plane wave excitation (50 QNMs were used). The extinction cross-section $\sigma_{\mathrm{ext}}$ was normalized to the geometric cross-section, which for the dielectric ring resonator is equal to $S=2R_{\mathrm{out}}h$.

{\bf Experimental studies}
The experiments were performed on $\mathrm{BaO}-\mathrm{Ln}_{2} \mathrm{O}_{3}-\mathrm{TiO}_{3}$ ceramic samples with a permittivity of $\varepsilon$ $\approx 80.05$ and a loss tangent of $\tan (\delta)=3.6 \cdot 10^{-4}$. The material used lends itself well to mechanical changes in geometry. The samples are centimeter-sized, each with an outer radius of $R_{\mathrm{out}}=16 \mathrm{~mm}$, which allows changing the $W / h$ parameter with high accuracy. First, ring resonators with specified $R_{\mathrm{in}} / R_{\mathrm{out}}$ ratios and a height corresponding to $W / h=0.42$ were made from cylinders, after that the height of each ring resonator was smoothly decreased. In total, 13 initial samples with different $R_{\mathrm{in}} / R_{\mathrm{out}}$ ratios in the range from 0.4 to 0.7 were used (Photographs of the samples are provided in Supplemental information, Sec.~IV). Each of them was changed in height an average of 15 times, which yielded a total of 203 ring resonators of different heights and radius ratios and, accordingly, 203 experimental spectra in the parametric space ($R_{\mathrm{in}} / R_{\mathrm{out}}, W / h$).

Far-field experiments were carried out in a microwave anechoic chamber within the range of $1-6~\mathrm{GHz}$~\cite{Bogdanov2019Jan, Rybin2015Dec}. For the far-field extinction measurements, the sample was placed on a custom mount, transparent in the microwave range, between two horn antennas at a distance of about $2$ meters from each other. One antenna generated a plane wave with a $\mathbf{k}$ vector along the axial axis of the ring resonator, and the other received the transmitted signal. The transmitting and receiving horns were connected to a vector network analyzer Rohde \& Schwarz ZVB20. This setup represents a two-port system that can be described by the scattering matrix $S$.

To measure the extinction cross-section, it is sufficient to obtain forward signals twice, with and without a sample. The extinction spectrum can be obtained using the optical theorem~\cite{larsson2013wideband}:
\begin{equation}
\sigma_\text{ext}=-\frac{4 \pi c}{\omega} \frac{L_{r} L_{l}}{L} \operatorname{Im}\left(\frac{S_{12}}{S_{12}^{b}}\right), \label{eq9}
\end{equation}
where $c$ is the speed of light, $\omega$ is the angular frequency, $L$ is the distance between antennas, $S_{12}$ is the measured transmission spectrum from the samples, and $S_{12}^{b}$ is the measured transmission spectrum in the absence of the sample.

\section*{Data availability}
The data that support the plots within this paper and other findings of this study are available from the corresponding author on reasonable request.

\section*{Code availability}
All the relevant computer codes supporting this study are available from the corresponding author upon reasonable request.

\section*{Acknowledgments}
The authors thank Pavel Belov and Yuri Kivshar for the discussion and Elizaveta Nenasheva (Ceramics Co. Ltd., St. Petersburg) for providing ceramic samples for measurements. The theoretical studies were supported by the Russian Science Foundation (Project 23-72-10059). The experimental studies were supported by the Russian Science Foundation (Project 23-12-00114). N.S. acknowledges support from the Advancement of Theoretical Physics and Mathematics “BASIS” Foundation.

\section*{Author contributions}
N.S. and F.Z. performed all experiments. N.S. performed all calculations and experimental data processing. M.B and N.S. studied the responses of quasinormal modes near singular points (q-BIC and EP). K.S. and M.S. described the mode coupling using temporal coupling mode theory. M.L., A.B., M.B., and N.S. analyzed the theoretical and experimental results and co-wrote the draft of the paper. M.L. served as the project supervisor. All authors discussed and contributed to the project.

\section*{Competing interests}
The authors declare no competing interests.

%\backmatter

\section*{Additional information}
\bmhead{Supplementary information} 

The online version contains supplementary information available at ....

\bmhead{Correspondence} Correspondence and requests for materials should be addressed to Andrey Bogdanov.

\end{document}

\title[Non-Hermitian Topology of a Subwavelength Resonator in Synthetic Space]{\ \ \ \ \ \ \ Supplementary Information for: \\Non-Hermitian Topology of a Subwavelength Resonator in Synthetic Space}

\author[1,2]{\fnm{N.} \sur{Solodovchenko}$^{\dagger\text{,}}$}

\author[1,2]{\fnm{F.} \sur{Zhang}$^{\dagger\text{,}}$}

\author[2]{\fnm{M.} \sur{Bochkarev}}

\author[2]{\fnm{K.} \sur{Samusev}}

\author*[1]{\fnm{M.} \sur{Song}}\email{kevinsmz@foxmail.com}

\author*[2]{\fnm{M.} \sur{Limonov}} \email{m.limonov@mail.ioffe.ru}

\author*[1,2]{\fnm{A.} \sur{Bogdanov}}\email{a.bogdanov@hrbu.edu.cn}

\affil[1]{\orgdiv{Qingdao Innovation and Development Center}, \orgname{Harbin Engineering University}, \orgaddress{\street{Qingdao}, \city{Shandong}, \postcode{266000}, \country{China}}}

\affil[2]{\orgdiv{School of Physics and Engineering}, \orgname{ITMO University}, \orgaddress{\street{Saint-Petersburg}, \postcode{197101}, \country{Russia}}}

%\affil[3]{\orgdiv{Ioffe Institute}, \orgname{St. Petersburg}, \orgaddress{\street{St. Petersburg}, \postcode{194021}, \country{Russia}}}

\affil[]{\textsuperscript{$\dagger$}These authors contributed equally to this work}

\maketitle

\tableofcontents

\newpage

\section{Derivation of the conditions for appearing the q-BIC and EP}
\begin{figure}[h]
\centering
\includegraphics[width=0.6\textwidth]{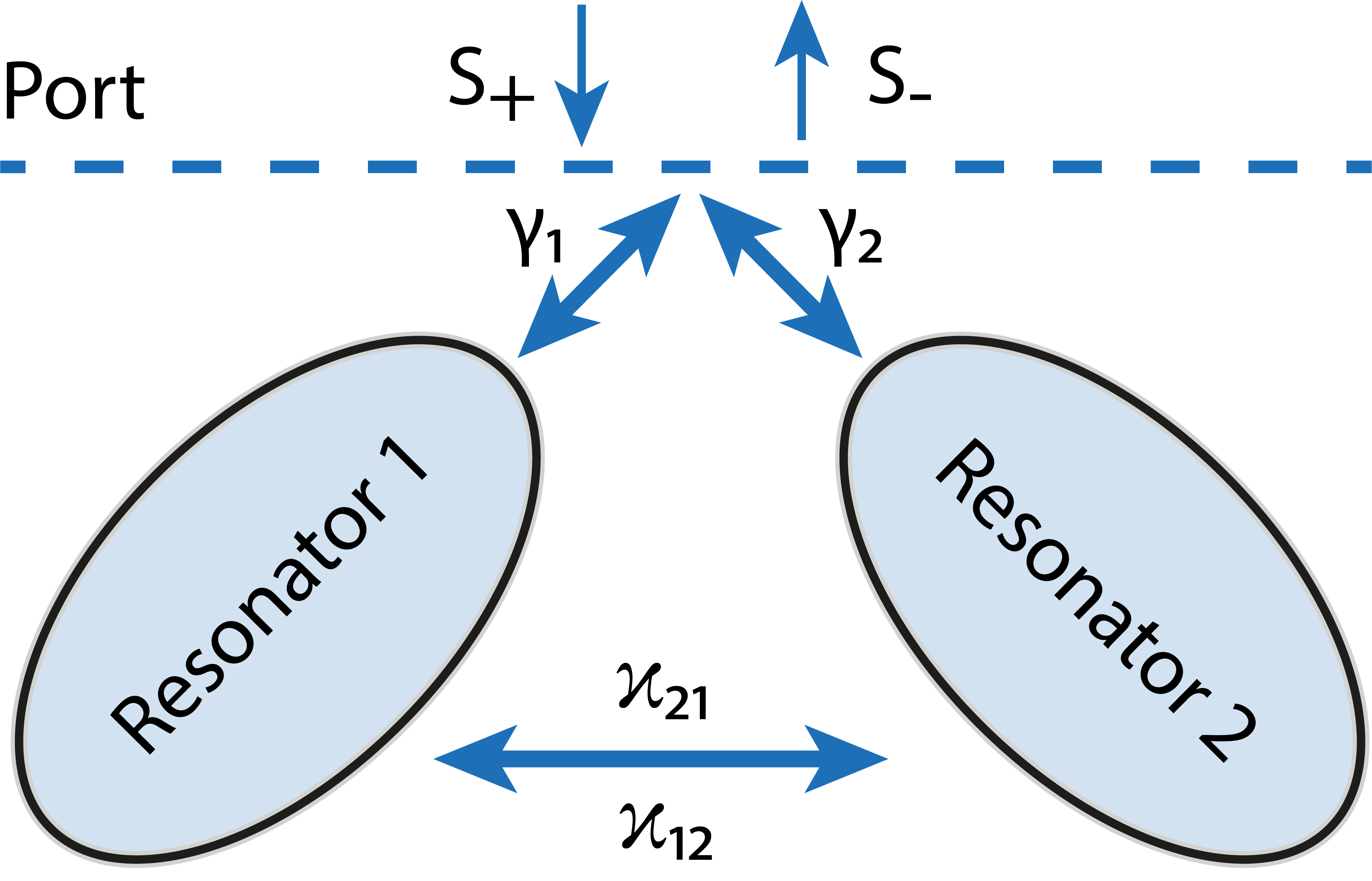}
\caption{$\mid$ A system of two resonators (or two eigenmodes) that couples to the same radiation channel.}\label{fig_S1}
\end{figure}

In this section, we derive the necessary conditions for the emergence of a q-BIC and an EP from a two-level Hamiltonian describing two resonances (eigenmodes) coupled directly through the near-field and via the same radiation channel (Fig.~\ref{fig_S1}).

Such a system can be described by the following Hamiltonian:

\begin{equation}
\mathcal{\hat H}=\left(\begin{array}{cc}\omega_{1} & \varkappa \\ \boldsymbol{\varkappa} & \omega_{2}\end{array}\right)-i\left(\begin{array}{cc}\gamma_{1} & \sqrt{\gamma_{1} \gamma_{2}} \\ \sqrt{\gamma_{1} \gamma_{2}} & \gamma_{2}\end{array}\right),\label{eq_S1} 
\end{equation}

\noindent \\ and characteristic equation for this Hamiltonian:
\begin{equation}
\widetilde{\omega}^{2}-\left[\left(\omega_{1}+\omega_{2}\right)+i\left(\gamma_{1}+\gamma_{2}\right)\right] \widetilde{\omega}+\left[\left(\omega_{1} \omega_{2}-\varkappa^{2}\right)+i\left(\gamma_{1} \omega_{2}+\gamma_{2} \omega_{1}-2 \varkappa \sqrt{\gamma_{1} \gamma_{2}}\right)\right]=0, \label{eq_S2}
\end{equation}
For that moment all notations corresponded to those from the main text, now we introduce new notations for simplicity: $\omega_{0}=\left(\omega_{1}+\omega_{2}\right) / 2 ; \Omega=\omega_{1}-\omega_{2} ; \gamma_{\mathrm{s}}=\gamma_{1}+\gamma_{2} ; \gamma_{\mathrm{d}}=\gamma_{1}-\gamma_{2} ; \gamma_{0}=$ $\sqrt{\gamma_{1} \gamma_{2}}$. Then, Eq.~\eqref{eq_S2} can be rewritten in the following form:
\begin{equation}
\left[\widetilde{\omega}-\left(\omega_{0}+i \frac{\gamma_{\mathrm{s}}}{2}\right)\right]^{2}=\frac{1}{4}\left[\left(\Omega^{2}+4 \varkappa^{2}-\gamma_{\mathrm{s}}^{2}\right)+2 i\left(\Omega \gamma_{\mathrm{d}}+4 \varkappa \gamma_{0}\right)\right], \label{eq_S3}
\end{equation}
with the solution for eigenfrequency:
\begin{equation}
\widetilde{\omega}=\left(\omega_{0}+i \frac{\gamma_{\mathrm{s}}}{2}\right) + \frac{1}{2} \sqrt{\left(\Omega^{2}+4 \mathcal{\varkappa}^{2}-\gamma_{\mathrm{s}}^{2}\right)+2 i\left(\Omega \gamma_{\mathrm{d}}+4 \varkappa \gamma_{0}\right)}. \label{eq_S4}
\end{equation}

It should be noted that the $"\pm"$ sign is not written in front of the radical, since in this case we are dealing not with an arithmetic root, but with the square root of a complex number of the form $a + i b$, where $a$ and $b$ are real numbers. The square root of such a number has two distinct values.
Let us examine this root in detail. To facilitate the analysis, we define these two key quantities:  
\begin{align*}
& a = \Omega^{2} + 4 \varkappa^{2} - \gamma_{\mathrm{s}}^{2} \\
& b = 2\left(\Omega \gamma_{\mathrm{d}} + 4 \varkappa \gamma_{0} \right).
\end{align*}
From fundamental complex analysis, it follows that:
$
r + i s = \sqrt{a + i b} \Rightarrow (r + i s)^{2} = a + i b.
$
By expanding this expression, we obtain 
$
a = r^{2} - s^{2} \ \text{and} \ b = 2 r s.
$
Next, substituting $ s = b/2r $ into the equation for $a$, we derive a biquadratic equation  for $r$
\begin{equation}
r^{2}-\frac{b^{2}}{4 r^{2}}-a=0 \Rightarrow r^{4}-a r^{2}-\frac{b^{2}}{4}=0, \label{eq_S5}
\end{equation}
This equation allows us to determine the possible values of $r$, which are crucial for understanding the system's behavior in the presence of coupled resonances. It's solution can be written as follows

\begin{equation}
\begin{split}    
& r_{ \pm}= \pm \frac{1}{\sqrt{2}} \sqrt{a+\sqrt{a^{2}+b^{2}}} \\
& s_{ \pm}=\frac{b}{2 r_{ \pm}}. 
\end{split}     \label{eq_S6} 
\end{equation}
Note that $r$ is a real number, and since $r^2$ cannot be negative, we select only one of the solutions from Eq.~\eqref{eq_S5} and utilize it to derive two distinct solutions in Eq.~\eqref{eq_S5}. As a result, we obtain the following two eigenfrequencies:

\begin{equation}
\widetilde{\omega}_{ \pm}=\left(\omega_{0} \pm \frac{1}{2 \sqrt{2}} \sqrt{a+\sqrt{a^{2}+b^{2}}}\right)+i \frac{1}{2}\left(\gamma_{s} \pm \frac{\sqrt{2} b}{\sqrt{a+\sqrt{a^{2}+b^{2}}}}\right). \label{eq_S7}
\end{equation}

%Now we will study behavior of the imaginary part $b/\sqrt{a+\sqrt{a^{2}+b^{2}}}$ with respect to $\Omega$ as a variable.
%We can find the roots for the derivative $$\partial\left(\frac{b}{\sqrt{a+\sqrt{a^2+b^2}}}\right) / \partial \Omega=0,$$ and obtain two corresponding solutions: $\Omega_{1}=\gamma \gamma_{\mathrm{d}}/\gamma_{0}$ and $\Omega_{2}=-\gamma_{0} \gamma_{\mathrm{d}}/\varkappa$. Substitution of the $\Omega_{1}$ in the derivative will provide an exact zero, and the $\widetilde{\omega}$ at this condition always have an extremum value in the imaginary part. Therefore, substitution of $\Omega_{1}$ in Eq.~\eqref{eq_S7} for $\omega_{\pm}$ correspond the q-BIC with coupling coefficient:

Next, we analyze the behavior of the imaginary part of $b/\sqrt{a+\sqrt{a^{2}+b^{2}}}$ as a function of $\Omega$. To determine the extrema, we solve for the roots of the derivative:
\[
\frac{\partial}{\partial \Omega} \left(\frac{b}{\sqrt{a+\sqrt{a^2+b^2}}} \right) = 0.
\]
This yields two solutions:

\[
\Omega_{1} = \frac{\gamma \gamma_{\mathrm{d}}}{\gamma_{0}}, \quad \Omega_{2} = -\frac{\gamma_{0} \gamma_{\mathrm{d}}}{\varkappa}.
\]
Substituting \(\Omega_{1}\) into the derivative results in an exact zero, indicating that \(\widetilde{\omega}\) always attains an extremum in its imaginary part under this condition. Therefore, substituting \(\Omega_{1}\) into Eq.~\eqref{eq_S7} for \(\omega_{\pm}\) corresponds to the formation of a q-BIC with the coupling coefficient (see Eq.~(2) in the main text):
\begin{equation}
{\varkappa}_{\text {BIC }}=\frac{\left(\omega_{1}-\omega_{2}\right) \sqrt{\gamma_{1} \gamma_{2}}}{\left(\gamma_{1}-\gamma_{2}\right)}. \label{eq_S8}
\end{equation}

%From the other side, we can study behavior of the real part $\sqrt{a+\sqrt{a^{2}+b^{2}}}$ with respect to $\Omega$. We can find roots for derivative $\partial\left(\sqrt{a+\sqrt{a^{2}+b^{2}}}\right) / \partial \Omega=0$ and discover that $\Omega_{2}=-\gamma_{0} \gamma_{\mathrm{d}}/\varkappa$ correspond to minimum value in real part of eigenfrequency, which corresponds to the Rabi point with: $\Omega_{\text {Rabi}}=\operatorname{Re}\left\{\widetilde{\omega}_{+}\right\}-\operatorname{Re}\left\{\widetilde{\omega}_{-}\right\}=\sqrt{4 \kappa^{2}-\gamma_{d}^{2}}$.

From the other side, we can examine the behavior of the real part 
$\sqrt{a+\sqrt{a^{2}+b^{2}}}$
as a function of $\Omega$. To determine the extrema, we solve for the roots of the derivative $\partial\left(\sqrt{a+\sqrt{a^{2}+b^{2}}}\right) / \partial \Omega=0$. It follows that $\Omega_{2} = -\frac{\gamma_{0} \gamma_{\mathrm{d}}}{\varkappa}$ corresponds to the minimum value of the real part of the eigenfrequency, which defines the avoid crossing point  with the following Rabi splitting
\[
\Omega_{\text{Rabi}} = \operatorname{Re} \left\{ \widetilde{\omega}_{+} \right\} - \operatorname{Re} \left\{ \widetilde{\omega}_{-} \right\} = \sqrt{4 \varkappa^{2} - \gamma_{\mathrm{d}}^{2}}.
\]

%Thus, at \(\Omega_{2}\), the system exhibits a minimal frequency splitting, indicating the Rabi oscillation regime.

From Eq.~\eqref{eq_S7} it is straightforward that conditions for the EP, correspond to $\sqrt{a+\sqrt{a^{2}+b^{2}}}=0$ and $\sqrt{2} b / \sqrt{a+\sqrt{a^{2}+b^{2}}}=0$, simultaneously; at the same time it is necessary to $a=0$ and $b=0$. Therefore, conditions for the EP:

\begin{equation}
\left\{\begin{array}{c}b = 2\left(\Omega \gamma_{\mathrm{d}}+4 \varkappa \gamma_{0}\right)=0 \Rightarrow \mathcal{\varkappa}_\text{EP}=-\frac{\left(\omega_{1}-\omega_{2}\right)\left(\gamma_{1}-\gamma_{2}\right)}{4 \sqrt{\gamma_{1} \gamma_{2}}},  \\ \Omega_{\mathrm{Rabi}}=\sqrt{4 \varkappa^{2}-\gamma_{d}^{2}}=0 \Rightarrow \mathcal{\varkappa}_\text{EP}= \pm\left|\gamma_{1}-\gamma_{2}\right| / 2 .\end{array}\right. \label{eq_S9}
\end{equation}
\\ When $|\mathcal{\varkappa}|<\left|\gamma_{1}-\gamma_{2}\right| / 2$ the weak coupling of modes and the intersection are observed (only in real or imaginary part), and $|\mathcal{\varkappa}|>\left|\gamma_{1}-\gamma_{2}\right| / 2$ represents the strong coupling and anticrossing. In the case of interaction of two modes in a single dielectric resonator, EPs always appear in pairs, which corresponds to different signs of $\varkappa_\text{EP}$. These two EPs are connected in parametric space by the bulk Fermi arc, along which real parts of eigenfrequencies are equal.

\begin{figure}[h]
\centering
\includegraphics[width=0.9\textwidth]{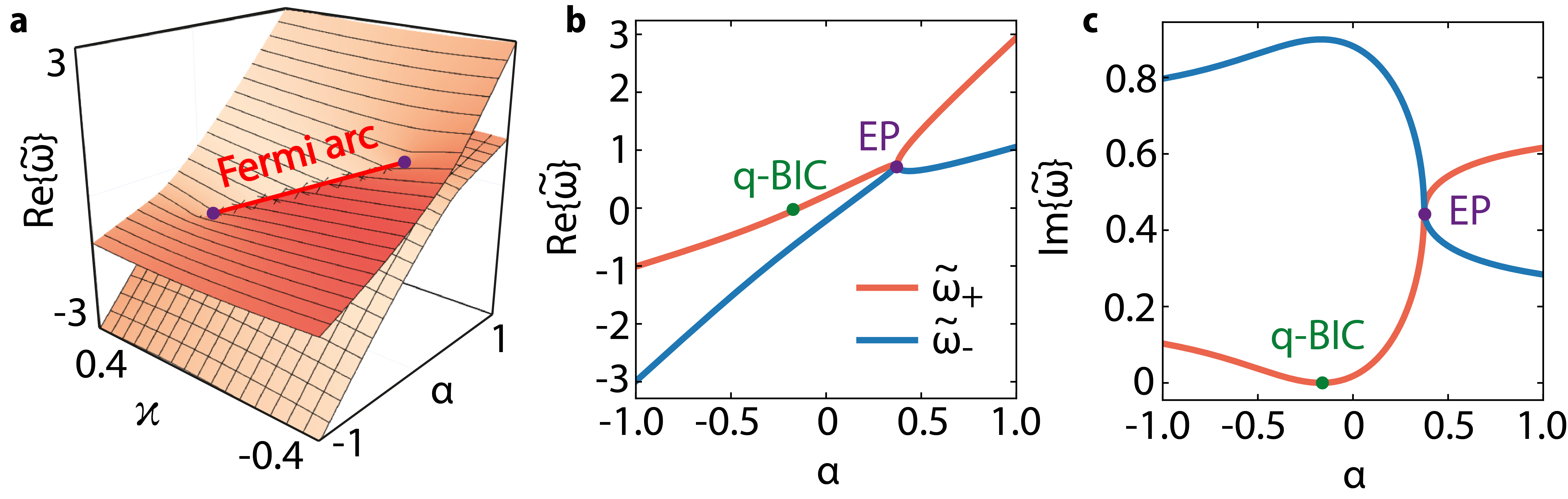}
\caption{$\mid$ {\bf Solutions for Hamiltonian} [Eq.~\eqref{eq_S1}]. {\bf a}, Eigenfrequencies $\widetilde{\omega}_{ \pm}$in parametric space and the Fermi arc (red line). {\bf b} and {\bf c}, Spectral positions of real and imaginary parts of eigenfrequencies $\widetilde{\omega}_{ \pm}$ with marked EP and BIC.}\label{fig_S2}
\end{figure}

For the demonstration, we plotted the results for $\widetilde{\omega}_{ \pm}$ for fixed $\gamma_{1}=0.7$ and $\gamma_{2}=0.2$ (Fig.~\ref{fig_S2}). The parameter $\alpha$ is related to frequencies of the uncoupled resonators in the next way: $\omega_{1}=$ $\alpha, \omega_{2}=3 \alpha$. We can consider $\alpha$ for example, as some geometric parameter of the resonator, the changes of which have different impact on uncoupled frequencies ($\omega_{1}$ and $\omega_{2}$). Such different impact can be observed for radial and axial modes in a RR upon changes in height - for radial modes impact on eigenfrequency is less than for axial (as demonstrated in the main text), thus such relation between $\alpha$ and ($\omega_{1}, \omega_{2}$) is a realistic simplification. Figure~\ref{fig_S2}{\bf a} shows behaviors of eigenfrequencies $\widetilde{\omega}_{ \pm}$ and the bulk Fermi Arc at which $|\mathcal{\varkappa}|<\left|\gamma_{1}-\gamma_{2}\right| / 2$ (blue line) and two EPs, where $\mathcal{\varkappa}_\text{EP}= \pm\left|\gamma_{1}-\gamma_{2}\right| / 2$. Figures~\ref{fig_S2}{\bf b} and ~\ref{fig_S2}{\bf c} show the real and imaginary parts of eigenfrequencies $\widetilde{\omega}_{ \pm}$at $\mathcal{\varkappa}=-0.25$, which corresponds to one of the EPs. When $\operatorname{Im}\left\{\widetilde{\omega}_{+}\right\} \rightarrow 0$, we observe BIC, which occurs at the condition from Eq.~\eqref{eq_S8}.

%Fig. S2 - Solutions for Hamiltonian (Eq. (S1)). (A) Eigenfrequencies $\widetilde{\omega}_{ \pm}$in parametric space and the Fermi arc (red line). (B) and (D) Spectral positions of real and imaginary parts of eigenfrequencies $\widetilde{\omega}_{ \pm}$with marked EP and q-BIC.

\begin{figure}[h]
\centering
\includegraphics[width=0.9\textwidth]{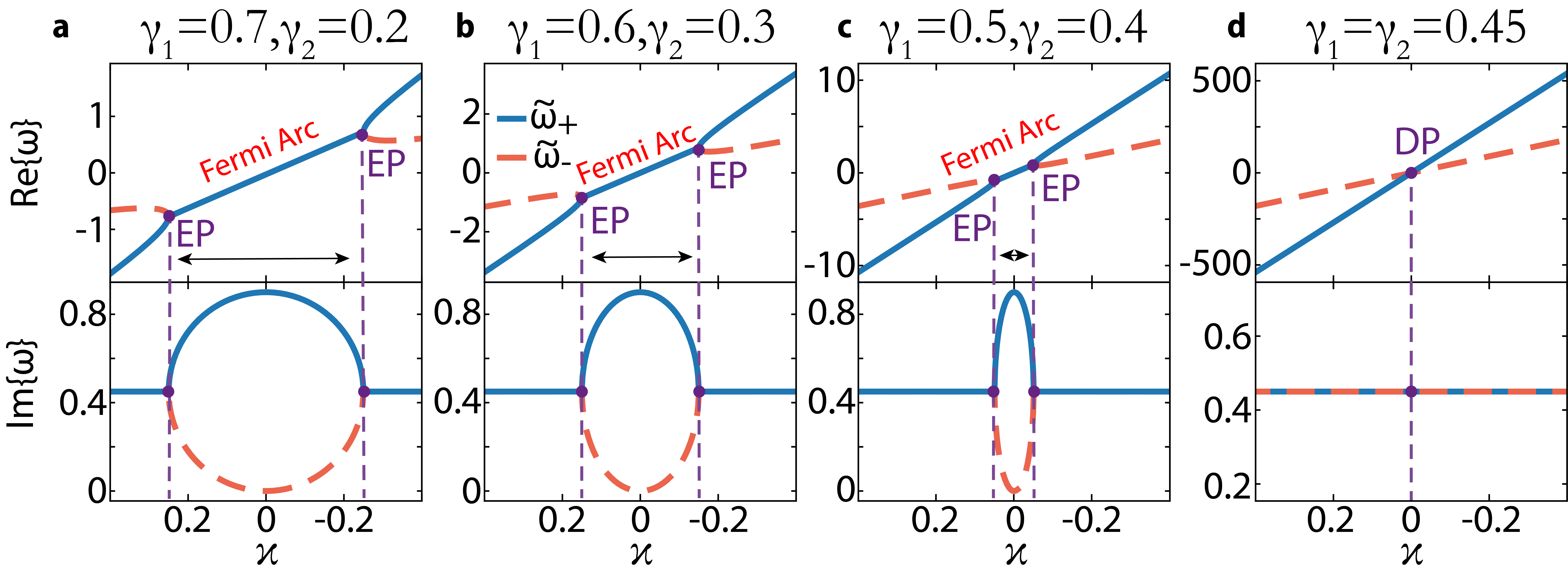}
\caption{$\mid$ {\bf Solutions for Hamiltonian} [Eq.~\eqref{eq_S1}] {\bf along Fermi arc with different values of resonance losses}. (A) Real and imaginary parts of eigenfrequencies $\widetilde{\omega}_{ \pm}$ with different values of resonance losses {\bf a} $\gamma_{1}=0.7, \gamma_{2}=0.2$, {\bf b} $\gamma_{1}=0.6, \gamma_{2}=0.3$, {\bf c} $\gamma_{1}=0.5, \gamma_{2}=0.4$ and {\bf d} $\gamma_{1} \approx \gamma_{2}=0.45$ marked above the figures.}\label{fig_S3}
\end{figure}

Equation~\eqref{eq_S9} shows that if $\left|\gamma_{1}-\gamma_{2}\right| / 2$ decreases, the Fermi arc becomes smaller and in the limiting case, when the losses of the two resonances are equal, degenerates into a Dirac point (DP). Figure~\ref{fig_S3} shows several examples for different values of resonance losses.

\begin{figure}[h]
\centering
\includegraphics[width=0.9\textwidth]{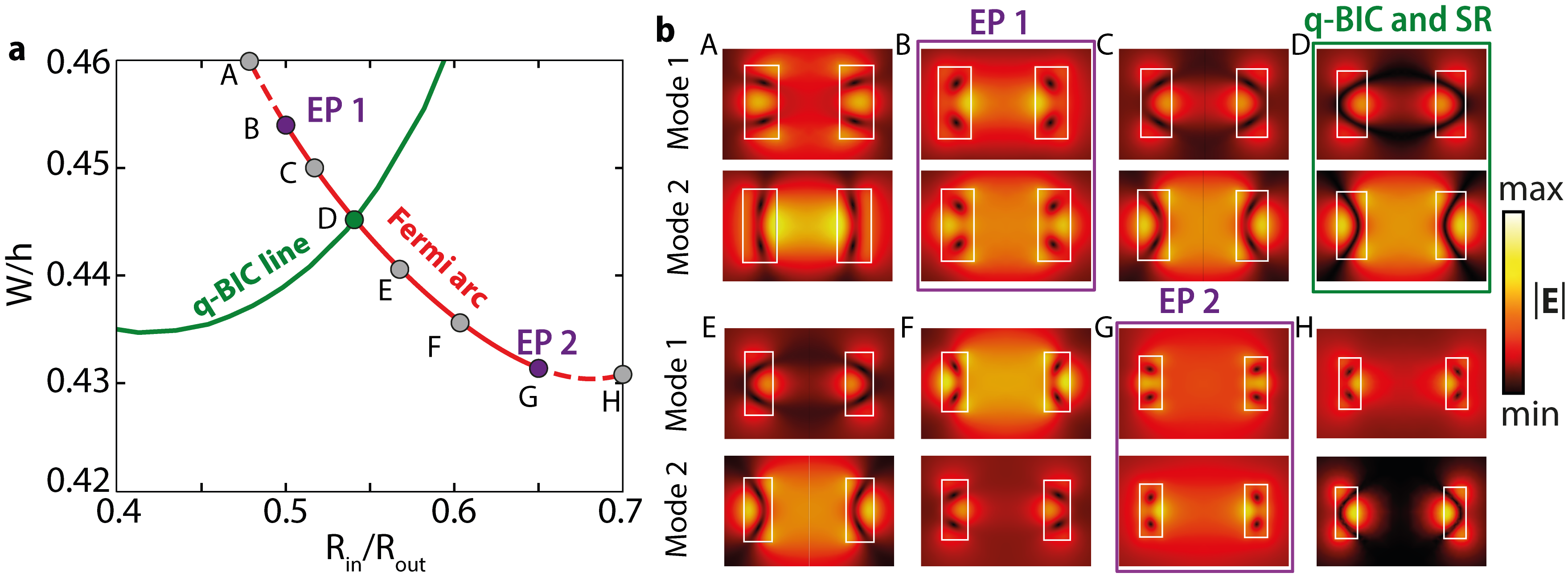}
\caption{$\mid$ {\bf The evolution of eigenmode fields along the Fermi arc}. {\bf a} Real part of eigenfrequencies in parametric space including the Fermi arc. {\bf b} Distributions of $|\mathbf{E}_{n}|$ in cross sectional plane for a number of different geometrical parameters, which correspond to the points in parametric space in {\bf a} (marked with letters from A to H)}.\label{fig_S4}
\end{figure}

Figure~\ref{fig_S4} shows a more detailed evolution of eigenmodes field distributions along the bulk Fermi arc for the real ring resonator that was shown in Fig. 2 in the main text. Along the bulk Fermi arc, not only 2 EPs can be detected, but also a BIC at the intersection of the bulk Fermi arc and the BIC line. Thus, BIC is observed on the Fermi arc.

%Fig. S3 - Solutions for Hamiltonian (Eq. (S1)) along Fermi arc with different values of resonance losses. (A) Real and imaginary parts of eigenfrequencies $\widetilde{\omega}_{ \pm}$with different values of resonance losses (A) $\gamma_{1}=0.7, \gamma_{2}=0.2$, (B) $\gamma_{1}=0.6, \gamma_{2}=0.3$, (C) $\gamma_{1}=0.5, \gamma_{2}=0.4$ and (D) $\gamma_{1} \approx \gamma_{2}=0.45$ marked above the figures.

\section{Bi-orthogonality, normalization and self-orthogonality of eigenmodes in the EP state}
Two eigenmodes are source-free solutions of the Maxwell equations $\mathcal{\hat H} \widetilde{\Psi}_{v, n}=\widetilde{\omega}_{v, n} \widetilde{\Psi}_{v, n}$, with $\widetilde{\Psi}_{n}=$ $(\widetilde{\mathbf{E}}_{n}, \widetilde{\mathbf{H}}_{n})$. Biorthogonality relation is based on the non-conjugate form of the Lorentz reciprocity theorem [1]:

\begin{equation}
\left(\widetilde{\omega}_{n}-\widetilde{\omega}_{v}\right) \iiint_{V} \widetilde{\Psi}_{n}^{T} \widehat{\mathbf{D}} \widetilde{\Psi}_{v}^{T} d^{3} \boldsymbol{r}=i \iint_{\Sigma}\left(\widetilde{\mathbf{E}}_{n} \times \widetilde{\mathbf{H}}_{v}-\widetilde{\mathbf{E}}_{v} \times \widetilde{\mathbf{H}}_{n}\right) \cdot d \boldsymbol{s}, \label{eq_S10}
\end{equation}
where  the integral in taken of over the entire volume $V$ including PML; $\Sigma$ is the integral over the PML surface, which is zero under PEC boundary conditions, and for the non-dispersive case $\widehat{\mathbf{D}}=\operatorname{diag}\left(-\mu_{0}, \varepsilon_{0}, \varepsilon_\text{res}\right)$, thus the right side of the equation will be zero. Thus, for $m \neq n$, the volume integral $\iiint_{V} \widetilde{\Psi}_{n}^{T} \widehat{\mathbf{D}} \widetilde{\Psi}_{v}^{T} d^{3} \boldsymbol{r}=0$, while for $m=n$ condition become $\iiint_{V} \widetilde{\Psi}_{n}^{T} \widehat{\mathbf{D}} \widetilde{\Psi}_{v}^{T} d^{3} \boldsymbol{r}=1$. The most general form of the PML norm for both dispersive and non-dispersive media can be obtained with the follwoing equation~\cite{lalanne:2018}:%[63]
\begin{equation}
\mathrm{QN}=\iiint_{V}\left[\mathbf{E}_{n} \cdot \frac{\partial (\omega \varepsilon)}{\partial \omega} \mathbf{E}_{n}-\mathbf{H}_{n} \frac{\partial (\omega \mu)}{\partial \omega} \mathbf{H}_{n}\right] d^{3} \boldsymbol{r}.
\end{equation}
The value of this integral for the non-normalized eigenmodes \(\left(\mathbf{E}_{n}, \mathbf{H}_{n}\right)\), calculated in COMSOL by varying the geometry parameter \( R_{in}/R_{out}, W/h\). Thus, the normalized field looks like this:

\[
\widetilde{\mathbf{E}}_{n} = \frac{\mathbf{E}_{n}}{\sqrt{\mathrm{QN}}}.
\]

Substituting \(\widetilde{\mathbf{E}}_{m}\) into Eq.~\eqref{eq_S5} in the main text results in a large value of \( \left|\xi_{n}\right| \), which corresponds to a giant intensity of a single quasinormal mode (QNM) in the vicinity of the EP. The tendency of the intensity can be attributed to the self-orthogonality of two nearly degenerate eigenvalues, both in terms of their eigenvalues, field distributions.

%================= вот тут дублируется с тем, что сверху=======
% The value of this integral for the non-normalized eigenmodes \(\left(\mathbf{E}_{n}, \mathbf{H}_{n}\right)\), calculated in COMSOL by varying the geometry parameter \( W/h \) while keeping the ratio \( {R}_{\mathrm{in}} / {R}_{\text {out }}=0.499 \) fixed, is shown in Fig.~\ref{fig_S4}. As the system approaches the exceptional point (EP), the integral value tends toward zero in modulus, leading to a significant increase in the normalized field:

% \[
% \widetilde{\mathbf{E}}_{n} = \frac{\mathbf{E}_{n}}{\sqrt{\mathrm{QN}}}.
% \]

% Substituting \(\widetilde{\mathbf{E}}_{m}\) into Eq.~\eqref{eq_S5} in the main text results in a large value of \( \left|\xi_{n}\right| \), which corresponds to a giant intensity of a single quasinormal mode (QNM) in the vicinity of the EP. The tendency of the integral \(\mathrm{QN}\) to approach zero can be attributed to the self-orthogonality of two nearly degenerate eigenvalues, both in terms of their eigenvalues and field distributions.
%=================

% \begin{figure}[h!]
% \centering
% \includegraphics[width=0.5\textwidth]{Figures/Fig_S5.png}
% \caption{$\mid$ {\bf PML-norm behavior near the EP}. $R_{in}/R_{out} = 0.4989.$}\label{fig_S5}
% \end{figure}

\begin{figure}[H]
\centering
\includegraphics[width=1\textwidth]{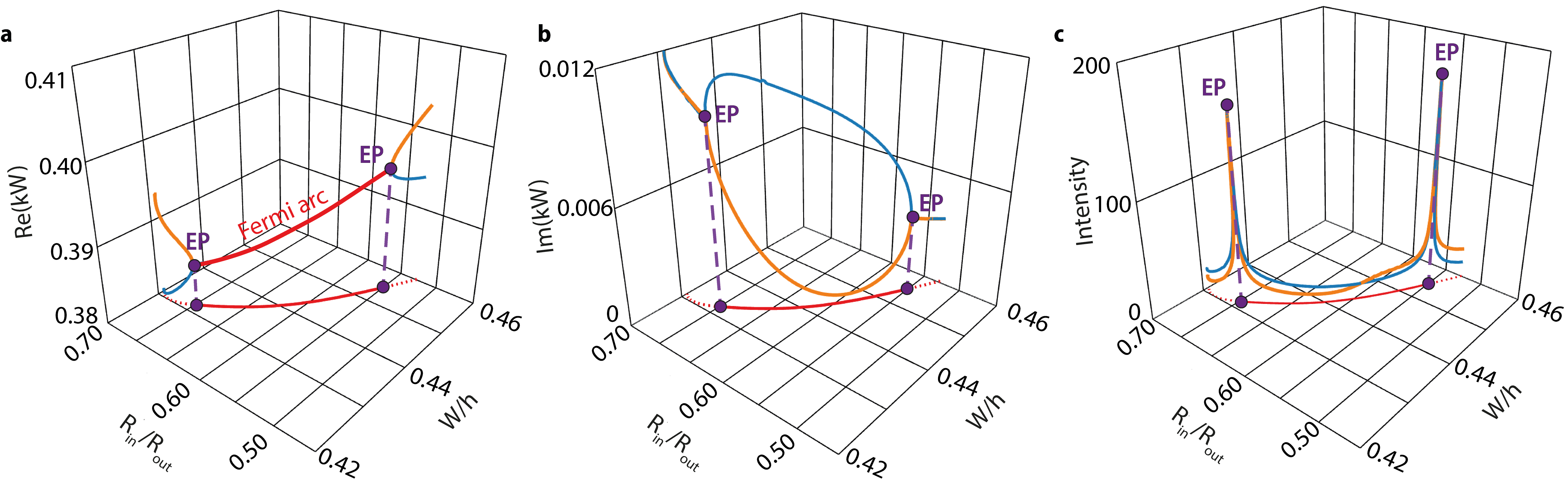}
\caption{$\mid$ {\bf The results of calculating the eigenvalues along the Fermi arc}: {\bf a} the real, {\bf b} imaginary part of the eigenvalues, and {\bf c} the corresponding extinction intensity of two QNM at normal incidence on the dielectric RR.}\label{fig_S5}
\end{figure}

The intensity of a single QNM is calculated with Eq.~\eqref{S15} along the Fermi arc, which was obtained by interpolating the intersection points of two QNM by a polynomial of the real or imaginary surface of the eigenvalues (Fig.~\ref{fig_S5}{\bf a}, and {\bf b}). Despite the fact that the intensities of the two QNM have large values (Fig.~\ref{fig_S5} {\bf c}), their total resonant response, as a result of destructive interference, will be weak in extinction spectra.

%Fig. $\mathbf{S 5}$ - The results of calculating the eigenvalues along the Fermi arc: the real, imaginary part of the eigenvalues and the corresponding extinction intensity of two QNM at normal incidence on the dielectric RR.

\section{\ Spectral parameters}

While the main text focuses on the reconstruction of the complex eigenvalue landscape and on the experimental identification of the bulk Fermi arc, here we emphasize an additional important point: the spectral coefficients entering the extinction line shape can be obtained analytically within the quasinormal-mode (QNM) formalism~\cite{lalanne:2018, bochkarev:2026}. %[63,64]

We consider the scattering response of the ring resonator under illumination by a linearly polarized plane wave incident along the symmetry axis $z$. In this geometry, only modes with azimuthal index $m=\pm1$ are excited. For such excitation, the extinction spectrum is naturally represented as a sum of partial modal contributions, each associated with a single QNM. In contrast to the total scattering cross-section, where several resonances may overlap within the same radiation channel, the extinction cross-section admits a mode-resolved decomposition and therefore provides direct access to the parameters of individual resonant states.

The partial extinction contribution of the $n^{th}$ QNM is written in the Fano form
\begin{equation}
\sigma_\mathrm{ext}^{n}=
\sigma_{n}
\left[
\frac{q_{n}^{2}-1+2q_{n}\Omega_{n}}
{(\Omega_{n}^{2}+1)(q_{n}^{2}+1)}
\right],
\label{S12}
\end{equation}
where $\Omega_n=(\omega-\omega_n)/\gamma_n$, and $\omega_n$ and $\gamma_n$ are the resonance frequency and half-width of the corresponding mode. The parameters $\sigma_n$ and $q_n$ characterize, respectively, the modal strength and the asymmetry of the Fano profile. The total extinction cross-section is then given by
\begin{equation}
\sigma_\mathrm{ext}=\sum_n \sigma_\mathrm{ext}^{n}+\sigma_\mathrm{bg},
\label{S13}
\end{equation}
where $\sigma_\mathrm{bg}$ is the nonresonant background term.

\begin{figure}[ht]
\centering
\includegraphics[width=0.8\textwidth]{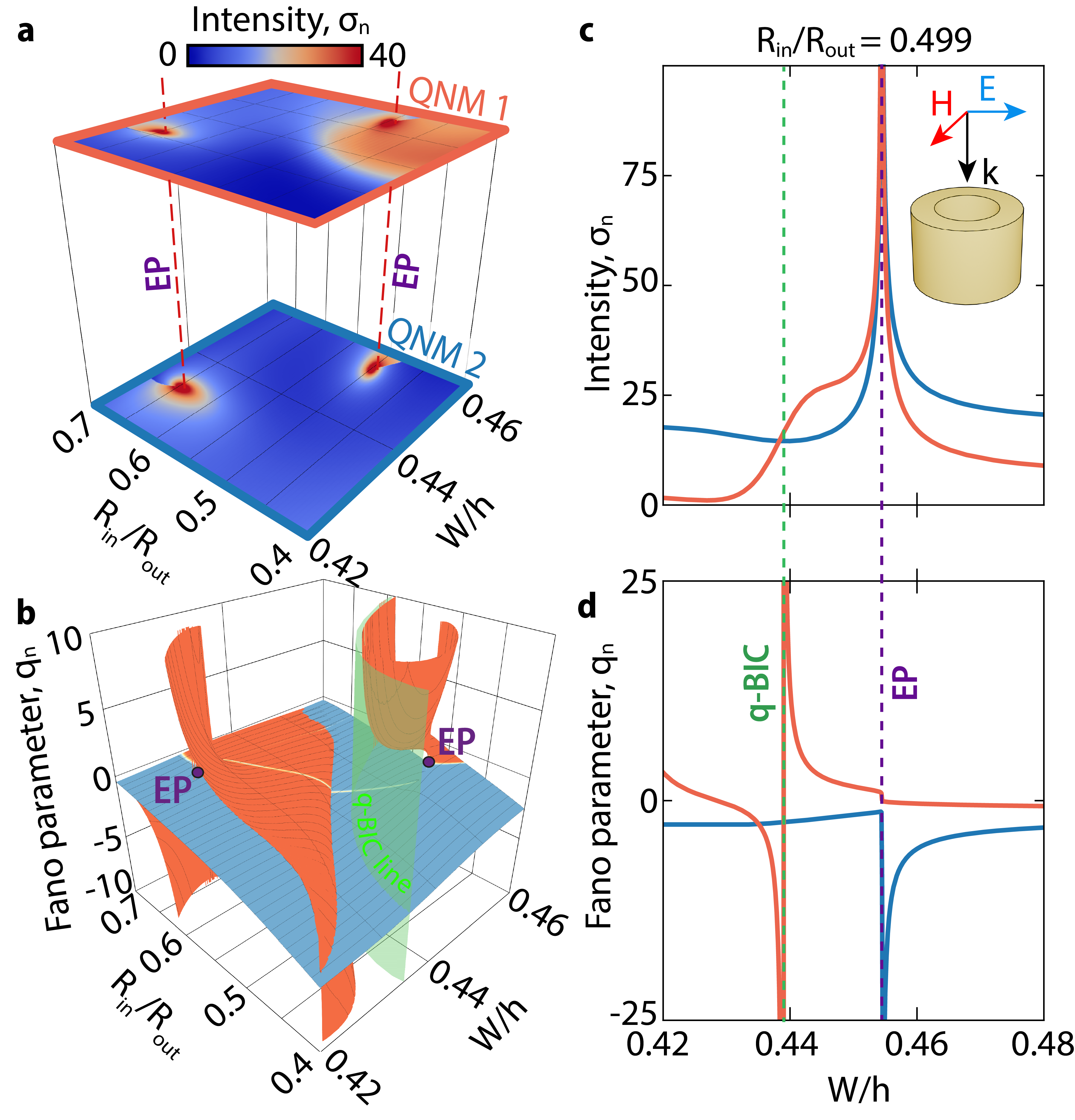}
\caption{$\mid$ {\bf Intensity and Fano parameter of EPs in parametric space at fixed excitation conditions.} {\bf a}, Behavior of intensities $\sigma_{n}$ for both eigenmodes near EPs in the parametric space ($R_{\mathrm{in}}/R_{\mathrm{out}}, W / h$), calculated using Eq.~\eqref{S15}. {\bf b}, Behavior of the Fano parameters $q_{n}$ for both eigenmodes near EPs in the parametric space ($R_{\mathrm{in}}/R_{\mathrm{out}}, W/h$), calculated using Eq.~\eqref{S14}. The green transparent surface corresponds to the q-BIC observation condition. {\bf c} and {\bf d}, slices of surfaces {\bf a} and {\bf b} along $R_{\mathrm{in}}/R_{\mathrm{out}}=0.499$, respectively, at which the first EP is observed. Excitation conditions: linearly polarized wave with $\mathbf{k}$ parallel to the symmetry axis of the RR. The inset in panel {\bf c} shows the incident wave.}\label{fig_S7}
\end{figure}

\newpage

A key advantage of the QNM approach is that all coefficients entering Eq.~\eqref{S12} can be calculated explicitly, without introducing fitting parameters. In particular, according to Ref.~\cite{bochkarev:2026}, %[64]
\begin{gather}
q_n=-\cot\left(\frac{\arg(\xi_n)}{2}\right), \label{S14}\\
\sigma_n=\frac{1}{P_0}\frac{\omega^2}{2\gamma_n}\left|\xi_n\right|, \label{S15}\\
\xi_n=
\int\limits_{V_{r'}} \Delta\varepsilon(\mathbf r')
\tilde{\mathbf E}_n(\mathbf r')\cdot
\mathbf E_b(\omega,\mathbf r')\,d^3\mathbf r'
\cdot
\int\limits_{V_r} \Delta\varepsilon(\mathbf r)\tilde{\mathbf E}_n(\mathbf r)\cdot
\mathbf E_b^*(\omega,\mathbf r)\,d^3\mathbf r.
\label{S16}
\end{gather}
Here $\tilde{\mathbf E}_n$ is the normalized QNM field, $\mathbf E_b$ is the background field (in our case, the incident plane wave), $P_0$ is the time-averaged Poynting vector of the incident wave, and $\Delta\varepsilon=\varepsilon_0(\varepsilon-\varepsilon_b)$ is the dielectric contrast between the resonator and the surrounding medium.

Thus, both the modal intensity $\sigma_n$ and the Fano asymmetry parameter $q_n$ are determined by overlap integrals between the excitation field and the corresponding QNM. Hence, the evolution of the extinction spectra in the parameter space $(R_\mathrm{in}/R_\mathrm{out},\, W/h)$ can be traced to analytically calculable modal characteristics rather than to a purely numerical fitting procedure.

Equations~\eqref{S14}--\eqref{S16} allow us to evaluate $\sigma_n$ and $q_n$ independently for both interacting modes across the entire parameter space; the corresponding maps are shown in Fig.~\ref{fig_S7}. In the vicinity of the EP, the intensities of the two modes increase sharply and become nearly equal. This behavior is associated with the onset of self-orthogonality, a well-known signature of an EP.

\begin{figure}[ht]
\centering
\includegraphics[width=0.6\textwidth]{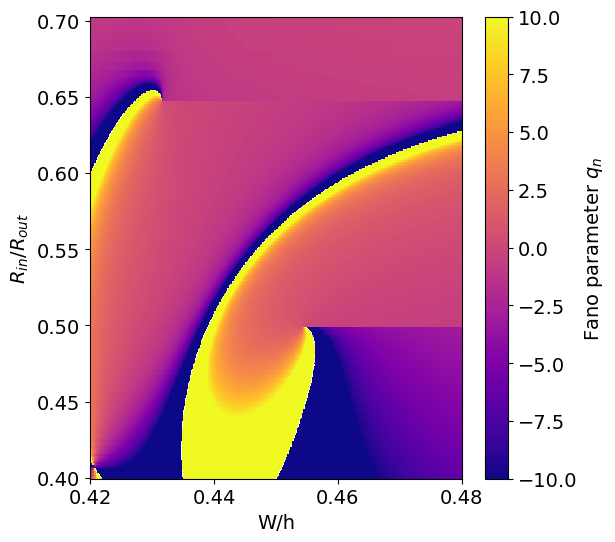}
\caption{$\mid$ {\bf Map of the Fano parameter of the EP in the parametric space under fixed excitation conditions.} Behavior of the Fano parameter $q_{n}$ for one eigenmode of eigenmodes near the EP in the parametric space ($R_{\mathrm{in}}/R_{\mathrm{out}}, W/h$), calculated using equation ~\eqref{S14}.}\label{fig_S8}
\end{figure}

The Fano parameter $q_n$ provides complementary information on the spectral line shape. Depending on the position in parameter space, the resonance evolves from an asymmetric Fano profile to an almost Lorentzian one at $q_n\to\pm\infty$, which in some cases corresponds to the q-BIC condition (see Fig.~\ref{fig_S7}b and Fig.~\ref{fig_S8}). Near the EP, the interacting modes may also satisfy $|q_1|=|q_2|$ with opposite signs, or $q_n=0$ for one mode and $q_n=\pm\infty$ for the other, corresponding to complete destructive interference and the formation of an anapole state~\cite{Zhang:2025}. %[57]%
Moreover, from Fig.~\ref{fig_S8}, one can see that the Fano parameter in our synthetic space twists in different directions - clockwise and counterclockwise, which indicates different topological charges of the two EPs. 

Therefore, the QNM framework not only reproduces the extinction spectra but also provides an analytical route to the key coefficients governing them. In this sense, the modal intensities and Fano asymmetry parameters serve as auxiliary indicators of the non-Hermitian singularities and can be evaluated independently for each mode throughout the entire parameter space.

\section{\ Fitting of spectra}
\begin{figure}[H]
\centering
\includegraphics[width=0.55\textwidth]{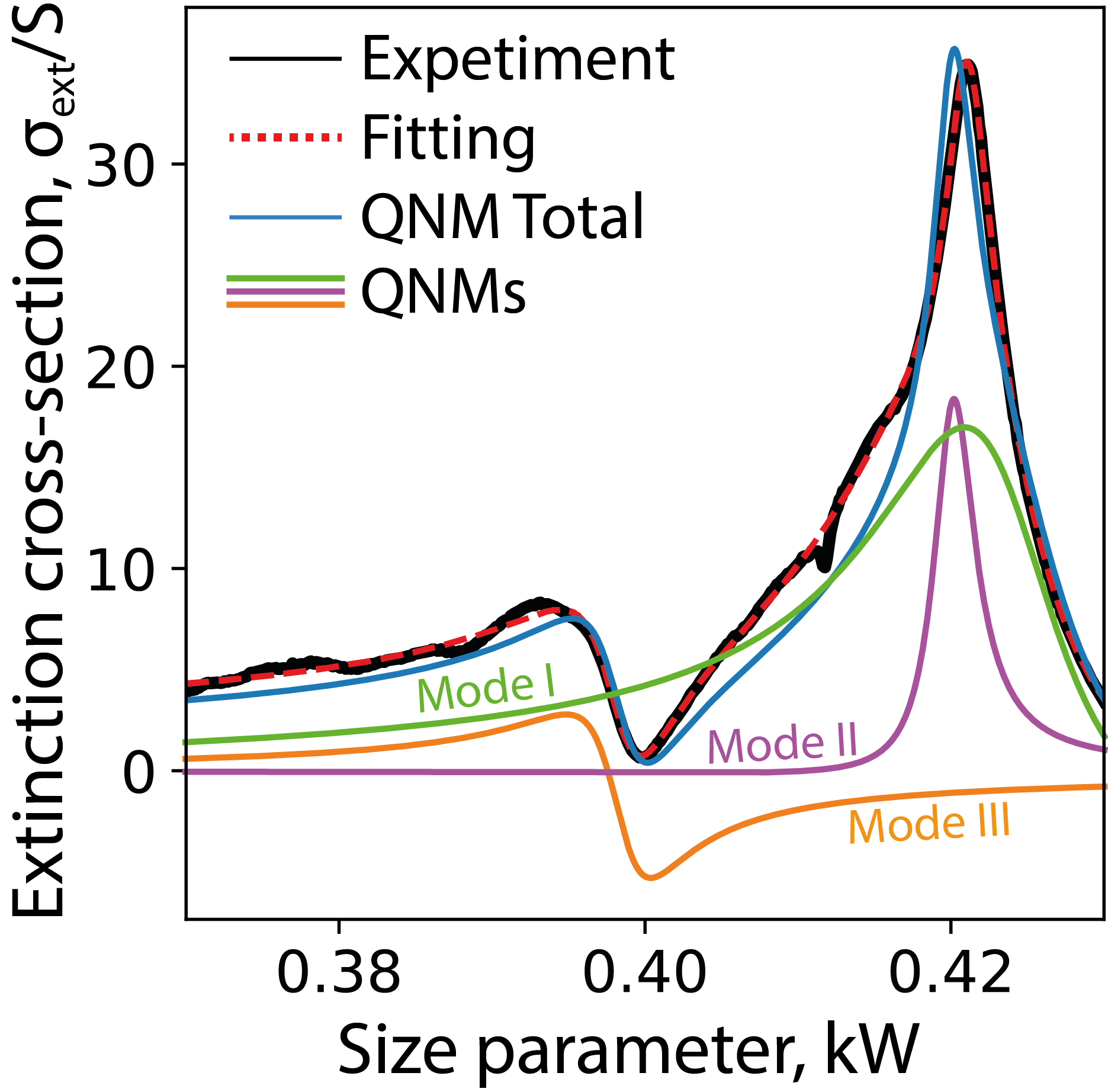}
\caption{$\mid$ Fitting of the experimental spectrum (black solid line) using 3 QNM parameters (orange, green and red solid line) as initial parameters $p_{0}=3 \times\left(\omega_{m}, \gamma_{m}, \sigma_{m}, q_{m}\right)$}\label{fig_S9}
\end{figure}

To fit the experimental spectra, the eigenvalues, intensity, and Fano parameters for a given incident wave for each experimental spectrum were first calculated for each experimental spectrum. The calculation showed that there are 3 eigenmodes in the frequency range of interest, which corresponds to 12 unknown parameters. Thus, for each experimental spectrum, 12 numerically calculated parameters were used as an initial approximation to find 12 experimental parameters. Figure~\ref{fig_S9} shows the experimental spectrum $R_{\text {in}}/R_{\text{out}}=0.5, W / h=0.494$, which was fitted (red dotted line) using parameters of 3 QNM (orange, green and purple solid line), the sum of which corresponds to the experimental spectrum.

%Fig. S6 - Fitting of the experimental spectrum (black solid line) using 3 QNM parameters (orange, green and red solid line) as initial parameters $p_{0}=3 \times\left(\omega_{m}, \gamma_{m}, \sigma_{m}, q_{m}\right)$.

\begin{figure}[ht!]
\centering
\includegraphics[width=0.9\textwidth]{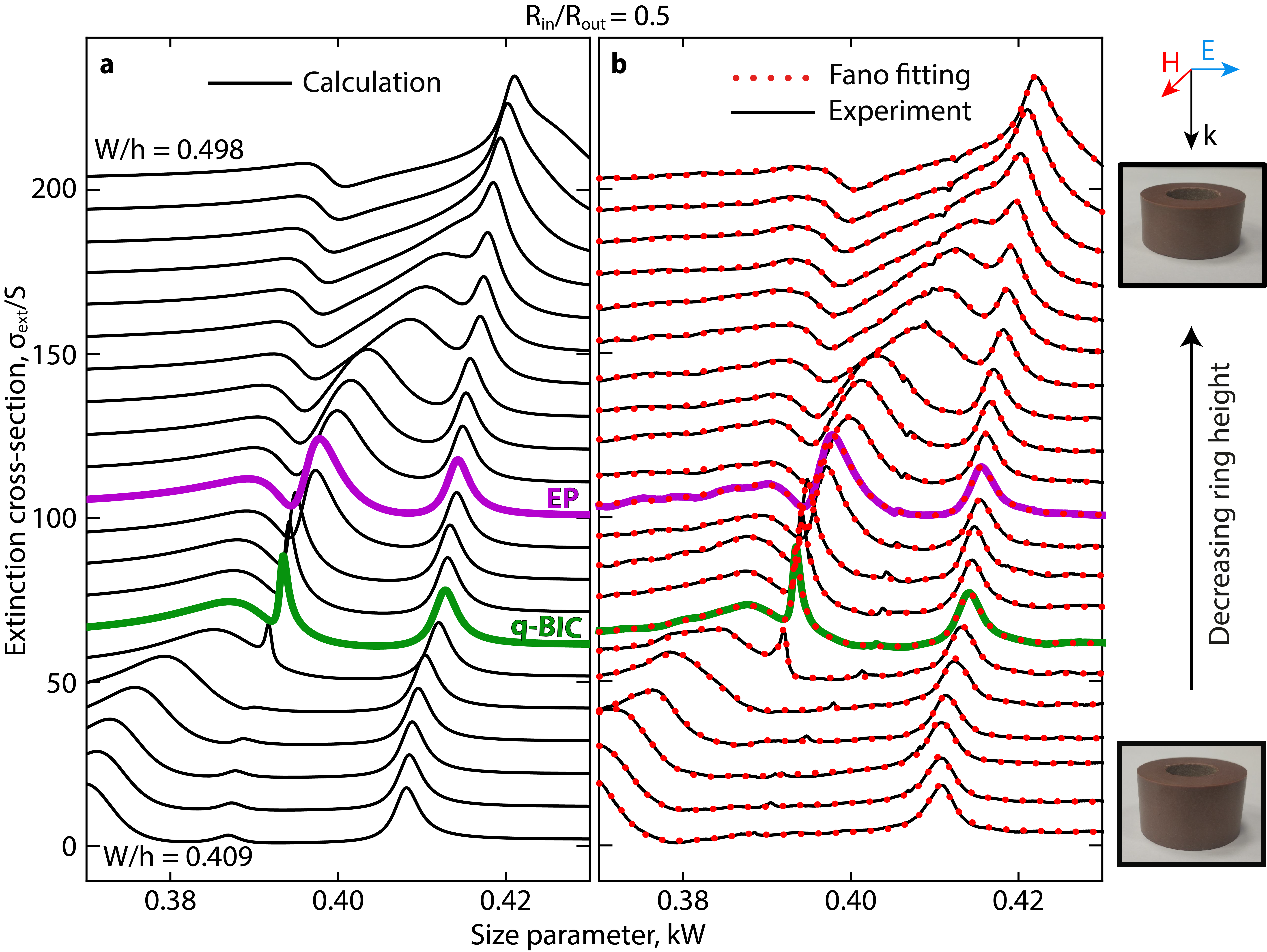}
\caption{$\mid$ {\bf Experimental extinction spectra of RRs, results of calculations and fitting.} {\bf a}, Calculated extinction cross-section spectra for fixed $R_{\mathrm{in}} / R_{\mathrm{out}}=0.50$ with variable ratio $W / h$. {\bf b}, Experimentally measured (black lines) and fitted (red dashed lines) extinction cross-section spectra for $R_{\mathrm{in}} / R_{\mathrm{out}}=0.50$. The spectrum corresponding to the q-BIC is highlighted with green color. The spectrum corresponding to the EP is highlighted with purple color. The excitation setup and the scheme of changing the RR height $h$ are shown on the right panel.}\label{fig_S10}
\end{figure}

As an example, Fig.~\ref{fig_S10}b demonstrates a series of experimentally measured extinction spectra for samples with variable height and a fixed ratio $R_{\mathrm{in}}/R_{\mathrm{out}}=0.50$, which is the closest parameter to EP obtained in numerical calculations for $R_{\mathrm{in}}/R_{\mathrm{out}} = 0.499$. The experimental spectra (Fig.~\ref{fig_S10}b) are in excellent agreement with the calculated ones (Fig.~\ref{fig_S10}a) for all 13 series of experiments, the width-to-height ratio parameters for which are presented in the Table~\ref{Table:Wh}. The red dotted lines in Figure~\ref{fig_S10}b correspond to the Fano fitting of the experimental spectra, which are in excellent agreement with the experimental curves.

\begin{table}[ht!]
\centering
\caption{Geometrical parameters $W/h$ for the case $R_{in}/R_{out}=0.5$, Fig.~\ref{fig_S10}. Spectrum numbers are given from top to bottom. The colors highlight the values at which EP (green) and BIC (purple) are observed, respectively.}
\label{tab:wh_rinrout_05}
\renewcommand{\arraystretch}{1.15}
\setlength{\tabcolsep}{6pt}

\begin{tabular}{c ccccccc}
\toprule
& \multicolumn{7}{c}{$R_{in}/R_{out}=0.5$} \\
\cmidrule(lr){2-8}
No. 
& 1 & 2 & 3 & 4 & 5 & 6 & 7 \\
$W/h$ 
& 0.498 & 0.493 & 0.488 & 0.484 & 0.479 & 0.475 & 0.472 \\
\midrule
No. 
& 8 & 9 & 10 & 11 & 12 & 13 & 14 \\
$W/h$ 
& 0.464 & 0.460 & 0.457 & {\color{epviolet}0.454} & 0.452 & 0.447 & 0.445 \\
\midrule
No. 
& 15 & 16 & 17 & 18 & 19 & 20 & 21 \\
$W/h$ 
& {\color{gbicgreen}0.443} & 0.435 & 0.426 & 0.420 & 0.414 & 0.411 & 0.409 \\
\bottomrule
\end{tabular}
\label{Table:Wh}
\end{table}

\newpage

\begin{figure}[ht!]
\centering
\includegraphics[width=0.45\textwidth]{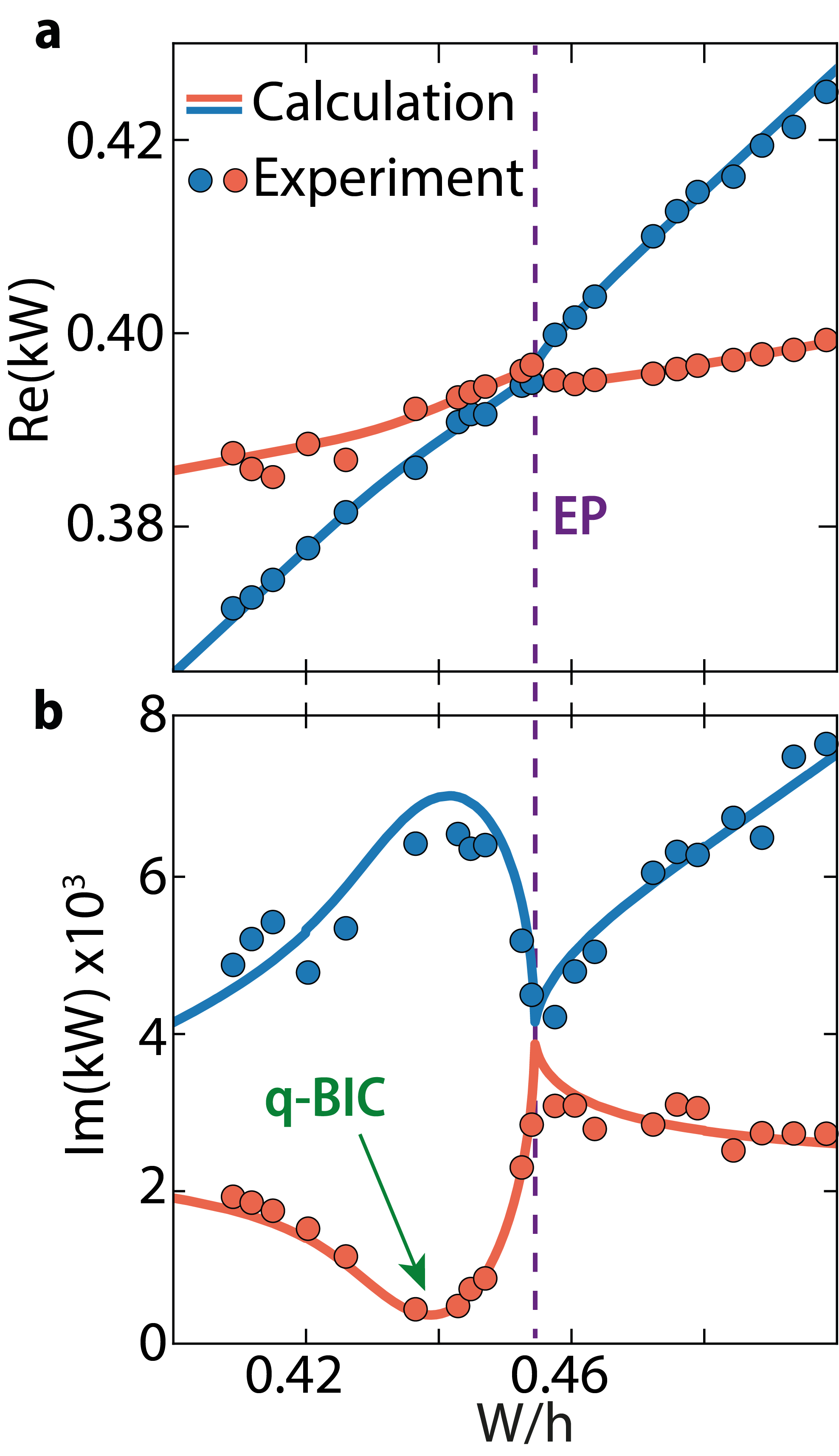}
\caption{\textbf{Experimental and numerical eigenfrequencies along $R_{\mathrm{in}}/R_{\mathrm{out}}=0.50$.}
(a,b) Real and imaginary parts of the dimensionless eigenfrequencies of the two interacting modes versus $W/h$. Solid curves are numerical results; circles are experimental eigenvalues reconstructed from Fano fits of the measured extinction spectra. The dashed vertical line marks the exceptional point (EP), whereas the q-BIC point is identified by the strong suppression of $\mathrm{Im}(kW)$ for one branch.}\label{fig_S11}
\end{figure}

Figure~\ref{fig_S11} shows a representative one-dimensional cut at $R_{\mathrm{in}}/R_{\mathrm{out}}=0.50$, corresponding to the experimental series closest to the numerically predicted exceptional point (EP). In this cut, the real parts of the two interacting branches approach each other near the singular point, whereas the imaginary parts reveal both the q-BIC condition and the subsequent eigenvalue coalescence at the EP. This representation provides a direct view of the spectral evolution within a single experimental series and illustrates how the non-Hermitian features emerge as the aspect ratio $W/h$ is varied.

\begin{figure}[ht!]
\centering
\includegraphics[width=1\textwidth]{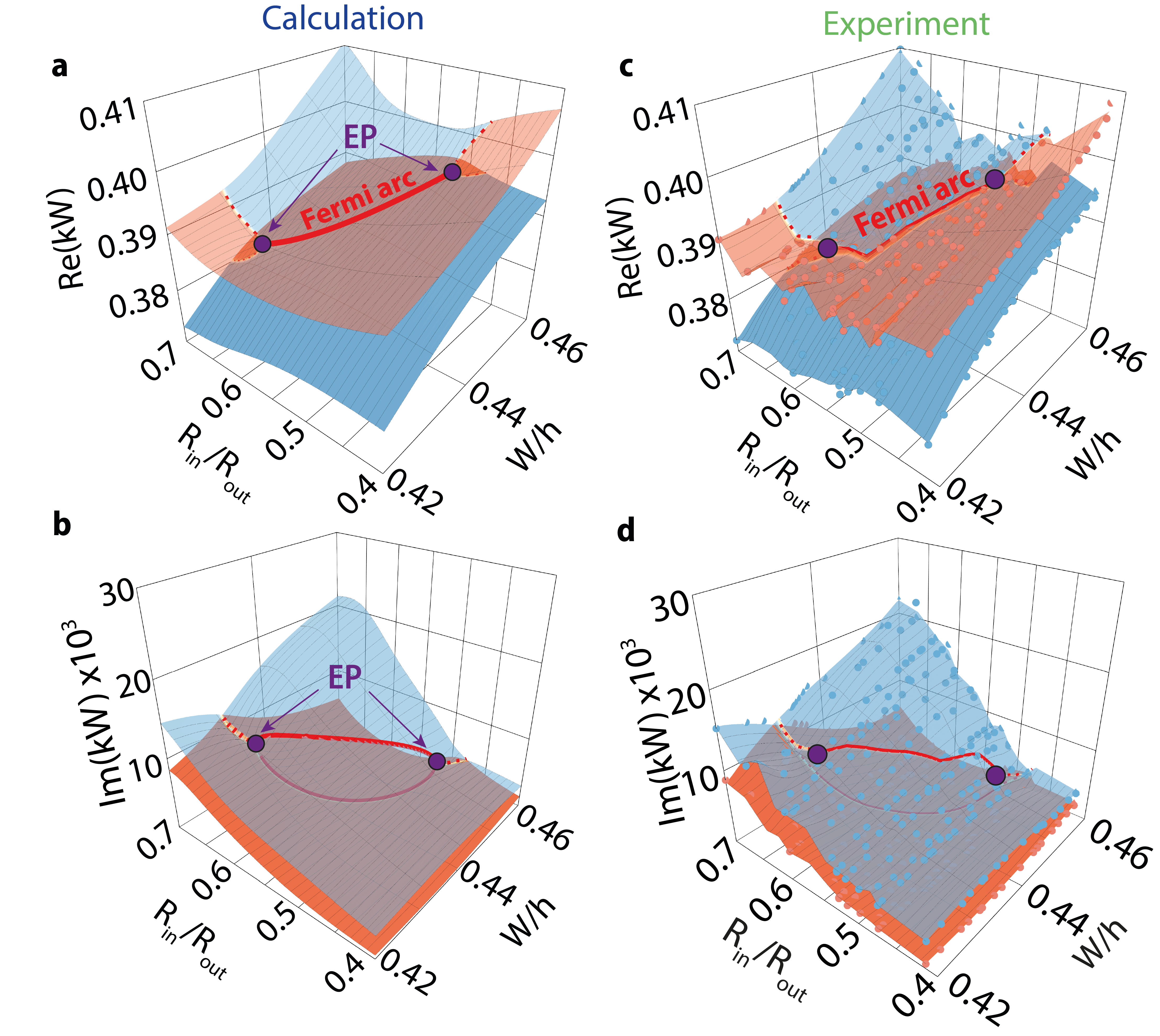}
\caption{\textbf{Numerical and experimental reconstruction of the complex-eigenvalue landscape in synthetic parameter space.}
(a,b) Numerically calculated real and imaginary parts of the dimensionless eigenfrequencies of the two interacting modes in the parameter space $(R_{\mathrm{in}}/R_{\mathrm{out}},\,W/h)$. 
(c,d) Experimentally reconstructed real and imaginary eigenvalue surfaces obtained by fitting the extinction spectra for all ring resonators shown in Fig.~\ref{fig_S16} and combining the extracted eigenfrequencies into a single dataset. The experimental reconstruction reproduces the topology of the numerical eigenvalue landscape, including two exceptional points (EPs) connected by a bulk Fermi arc.}\label{fig_S12}
\end{figure}

To reconstruct the full experimental eigenvalue landscape, the same procedure was repeated for each ring resonator shown in Fig.~\ref{fig_S16}. For every fixed value of $R_{\mathrm{in}}/R_{\mathrm{out}}$, the measured extinction spectra were fitted to recover the complex eigenfrequencies of the interacting modes as functions of $W/h$, after which all reconstructed eigenvalues were combined into a single dataset in the synthetic parameter space $(R_{\mathrm{in}}/R_{\mathrm{out}},\,W/h)$. Figure~\ref{fig_S12} presents the resulting two-dimensional reconstruction, which reproduces the topology of the numerical eigenvalue surfaces and demonstrates experimentally the presence of two EPs connected by a bulk Fermi arc.

\newpage

Examples of spectra fitting for each of the ring resonators are presented below. Figures~\ref{fig_S13}-\ref{fig_S15} demonstrate:

\begin{itemize}
  \item The first row: photographs of samples and parameter $R_{\mathrm{in}} / R_{\mathrm{out}}$.
  \item The second row is the experimentally measured (black lines) and fitted (red dashed lines) extinction cross-section spectra. The spectrum corresponding to the q-BIC is highlighted in green, the one corresponding to the EP is highlighted in purple.
  \item The third row and fourth rows are the dependencies of the real and imaginary parts of the dimensionless frequency of two RRs modes on the aspect ratio $W/h$, respectively.
\end{itemize}

\begin{figure}[H]
\centering
\includegraphics[width=1\textwidth]{Figures/Fig_S13.png}
\caption{$\mid$ Fitting of the experimental spectrum (black solid line) and eigenfrequencies vs W/h. Range of $0.4<R_{\mathrm{in}} / R_{\text {out}}<0.5$.}\label{fig_S13}
\end{figure}

%Fig. S7 - Fitting of the experimental spectrum (black solid line) and eigenfrequencies vs W/h. Range of $0.4<R_{\mathrm{in}} / R_{\text {out }}<0.5$.

\begin{figure}[H]
\centering
\includegraphics[width=1\textwidth]{Figures/Fig_S14.png}
\caption{$\mid$ Fitting of the experimental spectrum (black solid line) and eigenfrequencies vs W/h. Range of $0.5<R_{\mathrm{in}} / R_{\mathrm{out}}<0.57$.}\label{fig_S14}
\end{figure}

%Fig. S8 - Fitting of the experimental spectrum (black solid line) and eigenfrequencies vs W/h. Range of $0.5<R_{\text {in }} / R_{\text {out }}<0.57$.

\begin{figure}[H]
\centering
\includegraphics[width=1\textwidth]{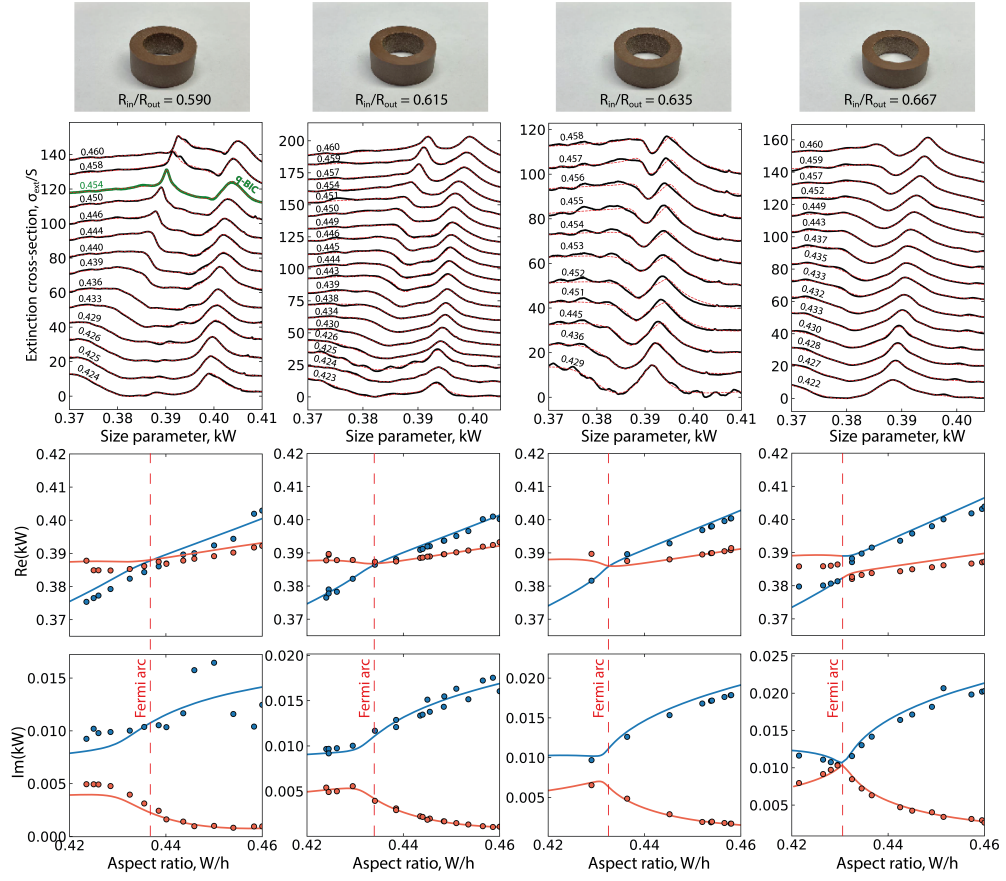}
\caption{$\mid$ Fitting of the experimental spectrum (black solid line) and eigenfrequencies vs W/h. Range of $0.58<R_{\mathrm{in}} / R_{\mathrm{out}}<0.7$.}\label{fig_S15}
\end{figure}

In each column corresponding to a fixed $R_{\mathrm{in}} / R_{\mathrm{out}}$, a photograph of the sample is presented at the top, and below are the experimental extinction spectra and their fitting (red dotted line) and the experimental eigenvalues (represented by dots) and a comparison with the calculations of the eigenfrequencies (solid lines). The extinction spectra corresponding to q-BIC are highlighted in green, and the closest spectra corresponding to EP are highlighted in purple. A combined photograph of all samples is shown in Figure~\ref{fig_S16}.

\begin{figure}[H]
\centering
\includegraphics[width=0.9\textwidth]{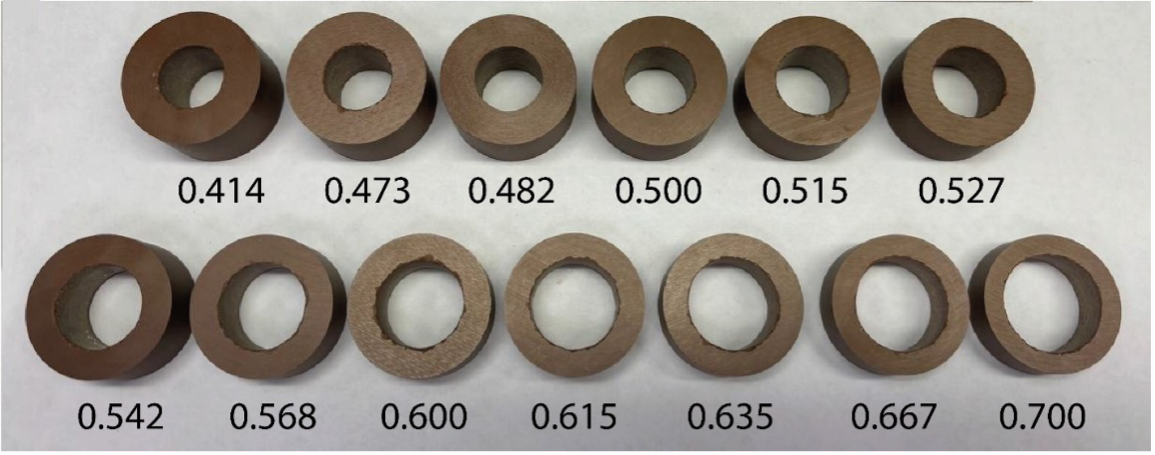}
\caption{$\mid$ Photograph of 13 RRs forming a two-dimensional  of 203 RRs used in the experiments. Changing the height of each RR allows for covering the entire parametric space ($R_{\mathrm{in}} / R_{\mathrm{out}}, W / h$) containing the bulk Fermi arc. The ratio of radii $R_{\mathrm{in}} / R_{\mathrm{out}}$ is indicated under each RR.}\label{fig_S16}
\end{figure}

\begin{figure}[H]
\centering
\includegraphics[width=0.75\textwidth]{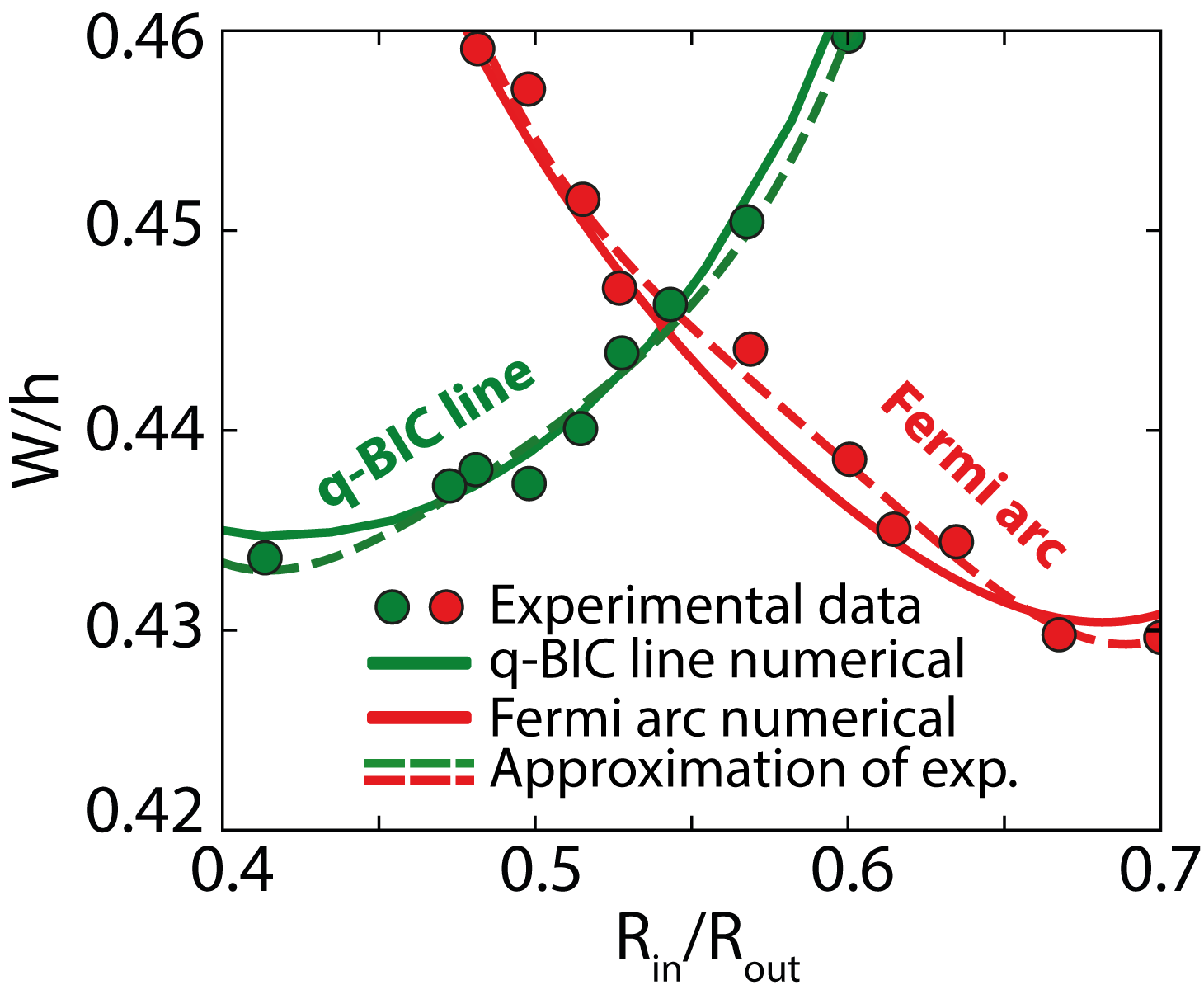}
\caption{$\mid$ Comparison of q-BIC and bulk Fermi arc lines obtained numerically and experimentally.}\label{fig_S17}
\end{figure}

Fitting each set of spectra for each of the presented samples allows us to determine the experimental position of the q-BIC and bulk Fermi arc. A comparison of the experimental bulk Fermi arc and the q-BIC line is shown in Figure~\ref{fig_S17}. The experimental points were approximated using a fourth-order polynomial. Thus, we have shown experimentally that the q-BIC and bulk Fermi arc intersect to form a topological robust state.

\section{\ Control of the bulk Fermi Arc}

\begin{figure}[H]
\centering
\includegraphics[width=1\textwidth]{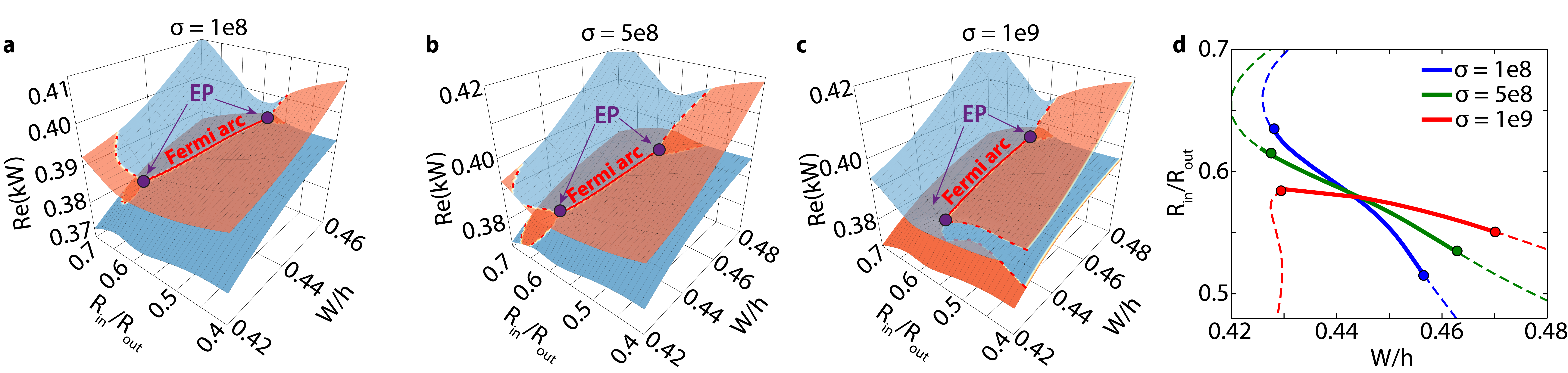}
\caption{Rotation of the bulk Fermi arc in the parametric space. {\bf a}-{\bf c} Riemann surfaces for ring resonators whose surface has an infinitely thin conducting layer (only surface is conductive, the bulk is dielectric) with conductance $\sigma = 10^8, 5\cdot10^8$ and $10^9$, respectively. The Fermi arc is highlighted with red color; all other parameters of the resonators are the same as for the Fig.2 from the main text. {\bf d} Rotation of the Fermi arc in parametric space with changes in conductivity.}\label{fig_S18}
\end{figure}

The bulk Fermi arc can be effectively controlled by introducing additional parameters into the system. For example, we have introduced an infinitely thin conducting layer on the surface of the ring resonator (all other parameters are the same as for Fig. 2 from the main text). The impact of such modification for a number of different values of the conductivity $\sigma$ are shown in Fig.~\ref{fig_S18}. Note that here we considered only the conducting surface; the volume of the resonator is still dielectric.    

The conductivity on the surface of the RR makes it possible to control radiation losses, thereby controlling the position of EPs and the bulk Fermi arc. Changes in the surface-conductivity produce peculiar behaviour of the bulk Fermi arc in parametric space, which rotates in counter-clockwise direction with increase of $\sigma$ (Fig.~\ref{fig_S18}d). Changing other parameters, for example, permittivity dependence on temperature, could open the way to new devices with dynamic optical properties.

